\documentclass[twocolumn,showpacs,amsmath,amssymb]{revtex4-2}
\usepackage{graphicx}
\begin{document}

\title{Elastic scattering at $\sqrt{s} = 6$ GeV up to $ \sqrt{s} = 13$ TeV \\
(proton-proton;  proton-antiproton; proton-neutron) } 


\author{  O.V. Selyugin\thanks{selugin@theor.jinr.ru} } 
\affiliation
{ BLTP, 
Joint Institute for Nuclear Research,
141980 Dubna, Moscow region, Russia } 
\begin{abstract}
 In the framework of the Regge-eikonal model of hadron interaction
 based on the analyticity of the scattering amplitude with taking into account the hadron structure,
   a simultaneous analysis  is carried out of  90 sets of data.
    These sets include the data were
    obtained at low energies ($\sqrt{s} > 3.6 $ GeV and  high energies
   at FNAL, ISR, $SP\bar{P}S$, TEVATRON and  LHC 
   with  4326 experimental points, including the polarization data of analysing power.
   The energy and momentum transfer dependence of  separate sets of data is analyzed on the basis of the eikonalized Born amplitude with taking into account  two additional anomalous terms.
  Different origins of the nonlinear behavior of the slope of the scattering amplitude are compared. 
 No contribution of hard-Pomeron in the elastic hadron scattering is   found. 
 The importance of Odderon's contribution is presented.
 The form and energy dependence of different terms of the cross-even and cross-odd parts of the elastic nucleon-nucleon scattering amplitude is determined in the framework of the High Energy Generalize Structure (HEGS) model. In the framework of the HEGS model, using the electromagnetic and gravitomagnetic form factors, the differential cross sections in the Coulomb Nuclear Interference (CNI) region and at large momentum transfer are described well in a wide energy region simultaneously.
  It is shown that the cross-even part includes the soft pomeron
 and the additional term with a large slope have  energy dependence $\ln^2(s)$.
  The cross-odd part includes the maximal odderon term  with $\ln^{2}(s)$    and an additional oscillation term with $\ln(s)$.
  It is shown that
   the additional terms with large slope are proportional to charge distributions but the maximal odderon term
    and oscillation term are proportional to  matter distributions.
 Also, a good description of proton-neutron differential scattering with 526 experimental points  is obtained  on the basis of the amplitudes taken from the analysis
  of $pp$ and $p\bar{p}$ scattering. A good  enough description of the polarization data was also obtained.
\end{abstract}
\maketitle

\section{Introduction}

  One of the most important tasks of modern physics is the research into the basic properties of   hadron interactions.
	The dynamics of strong interactions  finds its most
  complete representation in elastic scattering. 
  It is just this process that allows the verification of the
results obtained from the main principles of quantum field theory:
the concept of the scattering amplitude as a unified analytic function
of its kinematic variables connecting different reaction channels
was introduced in the dispersion theory by N.N. Bogoliubov\cite{Bogolyubov:1983gp}.
 Now many questions of  hadron interactions are connected with
 modern problems of  astrophysics such as unitarity and the optical theorem \cite{COTH-20},
 and problems of baryon-antibaryon symmetry and CP-invariance violation \cite{Uzik}.
  Many models predict that soft hadron interactions will enter a new regime at the LHC:
 given the huge energy, the S-matrix reaches the unitarity limit.
 The main domain of  elastic scattering is  small angles.
  Only in this region of interactions we can measure the basic properties that
  define the hadron structure. 
  Their values
  are connected, on the one hand, with the large-scale structure of hadrons and,
  on the other hand, with the first principles that lead to
  theorems on the behavior of scattering amplitudes at asymptotic
  energies \cite{mart,royt}.

    The  research of the structure of the elastic hadron scattering amplitude
   at superhigh energies and small momentum transfer - $t$
  can give a connection   between
  the  experimental knowledge and  the basic asymptotic theorems
    based on  first principles \cite{AKM,fp,royt}.
   This gives   information about the  hadron interaction
   at large distances where the perturbative QCD does not work \cite{PetOkor2,Drem1,Burk-15}
   and a new theory as, for example, instanton or string theories
    must be developed.

 The structure of  hadrons reflected in generalized parton distributions (GPDs)
   is now one of the most interesting questions                   
   of the physics of strong interactions (see, for example, \cite{Burk-N1,DD-N2}.
   This is tightly connected with the spin physics of the hadron \cite{Ter-gpd-sp}.
   Some modern accelerator experiments have developed extensive programs  for deep studies of
   different issues related to this problem.  For example, the Jefferson Laboratory/Electron-Ion Collider (EIC) team
   intends to extract  generalized parton distributions (GPDs) \cite{NPNews32}.
    The same problem is posed for
   future experiments at the SPD of NICA-JINR \cite{NICA}.

   Recent studies
   of elastic scattering of high energy protons lead to several unexpected results reviewed, e.g., in \cite{Drem1,Drem2}.
    Spin amplitudes of the elastic $NN$ scattering constitute a spin picture 
     of the nucleon.
   Without knowing of the spin $NN$-amplitudes, it is impossible to understand the spin observable
   of  nucleon scattering off nuclei.
       In the modern picture, the structure of hadrons is determined by  GPDs 
       which include the corresponding parton distributions (PDFs). The sum rules \cite{Mul94,Ji97a,Ji97b,R97}
     allow one to obtain the elastic form factor (electromagnetic and gravitomagnetic)
       through the first and second integration moments of GPDs \cite{R04}.  This leads to remarkable properties of GPDs,
     some   corresponding to inelastic  and elastic scattering of hadrons.
 Now  different
models examining the nonperturbative instanton contribution lead
to sufficiently large spin effects at superhigh energies \cite{KopZak,Forte,zpc,Dor}.
The research of such spin effects
will be a crucial stone for different models and will help us
to understand the interaction and structure of particles, especially at large
distances.
There are large programs of researching spin effects
at different  accelerators. Especially, we should like to note
the programs at the  NICA 
where the polarization of both the collider beams will be constructed.
So it is very important to obtain reliable predictions for  spin
asymmetries at these energies. In this paper, we extend the model
predictions to spin asymmetries in the NICA energy domain.

The unique experiment carried out by
     the  TOTEM Collaboration  at the LHC at 13 TeV gave  excellent experimental data
     on the elastic proton-proton scattering in a wide region of transfer momenta \cite{T66,T67}.
      It is  especially necessary to note the experimental data obtained at a small momentum transfer
     in the Coulomb-hadron interference region. The experiment reaches  very small
     $t = 8 \ 10^{-4}$ GeV$^2$  with small $\Delta t$, which give a large number of
     experimental points in a sufficiently small region of momentum transfer.
     This allows one to carry out careful analysis of the experimental data to explore
     some properties of  hadron elastic scattering.

   There are two sets of  data - at a small momentum transfer \cite{T66}
   and at a middle and large momentum transfer \cite{T67}.
   They overlap in some region of the momentum transfer, which supplies practically the same
   normalization of both sets of  differential cross sections
   of  elastic proton-proton scattering.
   Recently, the first set of data has created  a wide discussion of the determination of the total
   cross section and the value of $\rho(t=0)$  \cite{NM-rho}. 

    There is a very important  characteristic of the elastic scattering amplitude such as
    the ratio of the real part to imaginary part of the scattering amplitude - $\rho(s,t)$.
    It is tightly connected with the integral and differential dispersion relations.
    Of course, especially after different results obtained by the UA4  and UA4/2 Collaborations, 
     physicists understand that $\rho(s,t=0)$ is not a simple experimental value but heavily
    depends on  theoretical assumptions about the momentum dependence of the elastic scattering
    amplitude. Our analysis of both experimental data obtained by the UA4  and UA4/2 Collaborations
    shows a small difference  value of  $\rho(s,t=0)$  obtained in both  the experiments
       if the nonlinear
    behaviour of the elastic scattering amplitude is taken into account \cite{Sel-UA42}.
      Hence, this is  not an experimental problem
   but a theoretical one  \cite{CS-PRL}.

   For extraction of the sizes of $\sigma_{tot}$ and $\rho(t=0)$
   the Coulomb hadron region of momentum transfer is used (for example \cite{T66}). 
  However, the form of the scattering amplitude assumed for small $t$ and
  satisfying the existing experimental data
  at small momentum transfer can essentially be different from experimental data at large $t$.
   One should take into account the analysis of  the differential cross section
   at 13 TeV where the diffraction minimum impacts the form of $d\sigma/dt$  already at $t = -0.45$ GeV$^2$.

The analysis of  new effects discovered on the basis of the experimental  data at 13 TeV \cite{CS-PRL,Osc13,fd13}
  and associated with the specific properties of the  hadron potential
  at large distances was carried out
   with taking account all sets of experimental data on elastic $pp$-scattering obtained by
   the TOTEM and ATLAS Collaborations in a wide momentum transfer region
   and gave   a quantitative description of all examined experimental data with minimum
    fitting parameters.

    The non-linear behavior of the slope of the differential cross sections at a small momentum transfer,
    which was announced
   by the TOTEM Collaboration in the proton-proton elastic scattering at $8$ TeV, shows that
  the complex (complicated) form of  strong interactions searched out at low energies
   remains at superhigh   energies too.
   This  means that the strong interaction is not simplified at superhigh energies but
   includes many different parts of the hadron  potential. 

   Using the existing model of nucleon elastic scattering
   at high energies $\sqrt{s}= [3.6  - 14$] TeV  \cite{HEGS0,HEGS1}, which involves
   minimum of free parameters, we are going to develop its extended version aimed  to describe all available
   data on cross sections and spin-correlation parameters  at  lower energies down  to the  SPD NICA region.
   The model will be based on the usage  of known information on GPDs in the nucleon, electro-magnetic and gravitomagnetic  form factors of the nucleon
    taking into account  analyticity and unitarity requirements and providing compatibility with the high energy limit,  where the pomeron exchange dominates.

 The structure of the paper is as follows:
 in Section 2 
   the basic picture of  diffraction scattering is discussed  including 
    the nonlinear behavior of the slope of the differential cross section at small momentum transfer.
     In Section 3 the experimental data base used in the analysis is reviewed.
     In section 4 the asymptotic structure of the scattering amplitude is discussed shortly
     and the form factors used are particularly considered.
     Section 5 presents the different model approximations   used in the model
     and the structure of the scattering amplitude. 
     In section 6 the results of the fitting procedure and the behavior of the imaginary and real parts of the
     elastic scattering amplitude are considered.
     In section 7 the impact of the value of $\rho(s,t)$ and odderon contributions on the differential cross sections is analysed.
     Section 8 discusses  new effects  discovered during the model analysis of  experimental data. 
     In section 9 the obtained results for proton-neutron elastic scattering are presented.
      In section 10 the results of the model calculations of the spin correlation parameter $A_{N}(s,t)$ are present.
         Finally, the  obtained results  
          and some predictions
  of  our model are discussed in  Section 11.



\section{Small momentum transfer region}

\subsection{Electromagnetic scattering}

The $t$ dependence of the elastic peak is  one of the important issues, which
 can also  be related to optics.
 At high energy, elastic scattering at small angles has some similarity with light scattering in the
 Fresnel region.  The amplitude for the electric field propagating in the $z$ direction
  is given by the Rayleigh-Sommerfeld equation
\begin{eqnarray}
E(x,y,z)=-{i \over \lambda} \int   \int_{-\infty}^{+\infty}{ E(x',y',0) \frac{e^{ikr}}{r} \cos \theta}dx'dy'
\end{eqnarray}
 where $r=\sqrt{(x-x')^2+(y-y')^2+z^2}$,and $ \cos \theta = \frac{z}{r}$.
The integral can be performed analytically only for the simplest geometrical cases.
     At $x=kR >> 1$ it can be obtained   \cite{Newton}
 \begin{eqnarray}
         {\cal A}(\theta) \propto \frac{i J_{1}(x \sin\theta)}{ \sin\theta} \nonumber
       \label{sphera}
\end{eqnarray}
 Hence, even in the simplest case the elastic peak cannot be described by a simple exponential.
 Note that this form of the scattering amplitude is used in  diffractive hadron scattering.

\subsection{Hadron scattering}

 In ref. \cite{TOTEM-8nexp}, the TOTEM Collaboration announced the observation of the non-exponential behavior of the differential
elastic cross sections at $8$ TeV and small momentum transfer $|t|$.
Of course, the form of the elastic peak depends also on the structure of  particles and  the dynamics of the interaction.
 These two features can be parameterised by the profile function $\rho(\vec b)$ (in the space of impact parameter, $\vec b$)
 or by the elastic form factor (in momentum space).
 For example, in \cite{Alberi-81}  different forms of the profile function determined by the
    density distributions taking part in the interaction were considered:
\begin{eqnarray}
 circle\ of\ radius\  b_0=4 \ GeV^{-1}   
\nonumber \\
 \Rightarrow     {\cal A}(t)  \propto  J_{0}( 4 \sqrt{|t|}); \label{F2}  \\
hollow\ disk\ near\  b_0=2.8 \ GeV^{-1} 
 \nonumber \\
\Rightarrow
       {\cal A}(t)  \propto  e^{2.8 t} J_{0}(2.8 \sqrt{|t|}); \label{F4}  \\
black \ disk\ of \ radius\ 6 \  GeV^{-1} \nonumber \\
\Rightarrow 
         {\cal A}(t)  \propto   J_{1}( 6 \sqrt{|t|} )/(6 \sqrt{|t|}); \label{F3}\\
\rho(b)\sim e^{-cb} 
\Rightarrow 
         {\cal A}(t)  \propto  e^{5 t}; \label{F1}   \\
\rho(b)\sim e^{ \mu \sqrt{b_{0}^2+b} }  
\Rightarrow
  {\cal A}(t)  \propto  e^{5 (\sqrt{4\mu^2-t}-2\mu )}. \label{F5}
\end{eqnarray}

\begin{figure}[h]
\begin{center}
\includegraphics[width=0.45\textwidth] {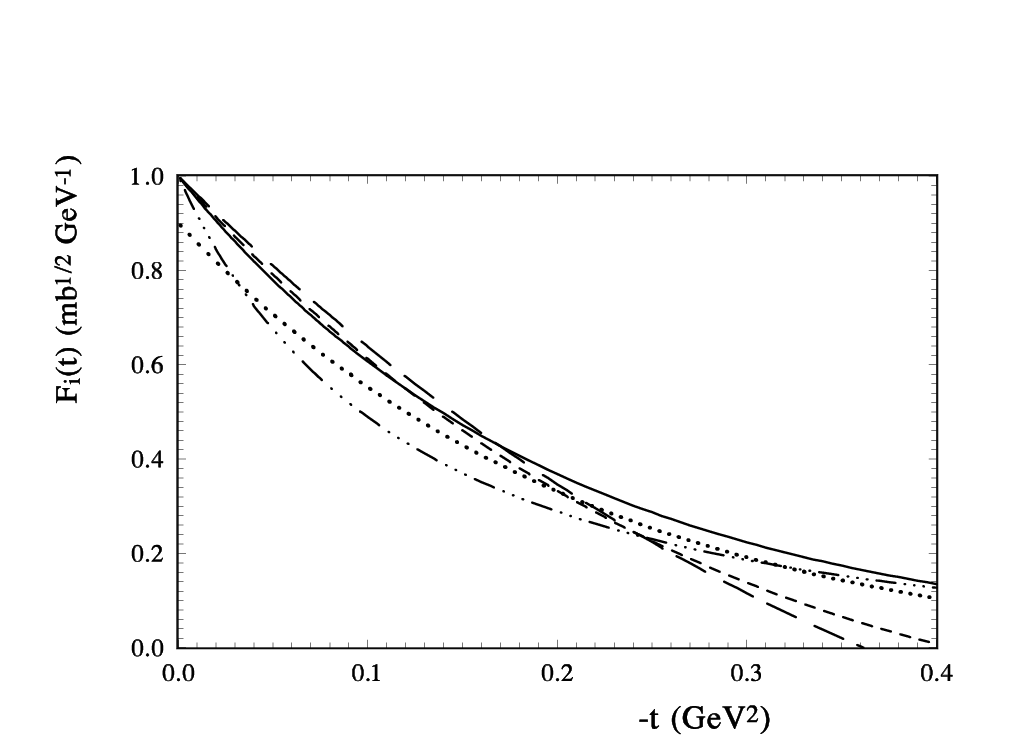}
\end{center}
\caption{ Possible $t$ dependences of the scattering amplitude ${\cal A}(t)$, rescaled to unity at $t=0$.
 (solid line - for Eq. (\ref{F1});  long-dashed line - for Eq. (\ref{F2});
  short-dashed line - for Eq. (\ref{F3}); dotted line  - for Eq. (\ref{F4}),
  and dashed-dotted line  - for Eq. (\ref{F5}).}
\end{figure}

The $t$ dependence of these amplitudes is shown in Fig. 1. Clearly, the simple exponential form (\ref{F1})
 is a special separate case.
 We can see, except for the last case (eq.6), that if we take a small interval of $t$ (for example,
  $0.01< |t|
  <0.15$ GeV$^2$) the curves  give practically the same result,  especially the lines of eqs. 3,5,6.
  However, they give essentially different results
 on some  larger values of $t$ or as $t \rightarrow 0$.
 Hence, to obtain the true form of the scattering amplitude, it is necessary to take
    experimental data in a wider interval of $t$ and tending to the limit $t \rightarrow 0$.

     This simplest approximation of the differential cross section by one exponent with a constant slope can lead
     to  artificial new effects. For example, in the experiment at the Protvino accelerator on elastic proton-proton
     scattering
     there were found  "oscillations" at a small momentum transfer \cite{Protv-osc}. However, in \cite{Sel-Prot} it was shown that
     such "oscillations" can appear when the differential cross section
     is described by two exponentials with different slopes and  one exponential with the constant slope.

 The investigation of the energy and $t$ dependence of the slope of the elastic peak can lead to new information about  the structure of  interacting particles.
 Early measurements at ISR 
  revealed four slopes of  differential cross section in  different regions of momentum transfer,
   which result from the complicated structure of  nucleons.

\subsection{Non-linear slope}

   The  complicated $t$ dependence of the slope can have
    many origins. First of all, it comes from the unitarization procedure of the Born
   term of the elastic scattering amplitude.
   In many purely phenomenological analyses it is
   represented by the $C t^2$ term in the slope that mimics the unitarization procedure.
       The hadron structure is  reflected also not only at a large momentum transfer but also
       at small  $t$.
       This especially  concerns the meson clouds whose interaction in many
       models is added to the central part of hadrons (for example,
       in the Pumplin model \cite{Pumplin-92} and the Dubna Dynamical (DD) model \cite{DDM}
       which leads to the
       $\sqrt{t_{0} +t}$
       dependence of the scattering amplitude).
    Such a complicated hadron structure can be reflected in the presence of two form factors
    - electromagnetic and mater form factor of hadrons (for example, the HEGS model \cite{HEGS0,HEGS1}).

     Another term of the slope, which is commonly represented as
     \begin{eqnarray}
    \sqrt{4 \mu^2 -t} -2 \mu,
    \label{stsq}
    \end{eqnarray}
    is used in many phenomenological descriptions of the elastic differential cross sections
      to explain the "break"   in the differential cross sections.
    Note that an additional term in the slope like
    $\sqrt{t_{0} - t}$
    was    obtained earlier   whose first approximation can be related to absorbtion corrections
      that can produce a set of canceling Regge cuts \cite{ChengWu-70,Cardy-71}
     and leads to the  Schwarz type trajectories \cite{Schwarz-68}
     $\alpha(t)=1+ \gamma t^{1/2}$.
     A more complicated form was obtained in \cite{Bronzan} 
      $\alpha_{\pm}(t)=1 \pm \gamma t^{1/2} + 2 \rho(1/2 \ \gamma^{2}t)^{3/2} (-ln(t))^{1/2}$.
     The  appearance of a complex trajectory greatly complicates the  picture and
     requires  additional research.
      Hence, in \cite{Jenk-72}, based on the works \cite{Barut-62,Barut-63}, it
       was proposed to use the simplest form $\alpha(t)=1.041-0.15 \sqrt{t_{0}-t}$.
       However, as we show, such behavior has a really pure phenomenological basis and can be
       replaced by  either a simpler form (for example, eq.(7) 
       or a more complicated form
       used in the HEGS-model \cite{HEGS1}).

   Many models based on the famous works
   \cite{Gribov-61,G-P-62} 
    researched the  non-linear behavior of the scattering amplitude.
Based on the works
in \cite{Gribov-Sl72}, it was obtained that
  \begin{eqnarray}
   \alpha_{P}(q^{2}) = 1-C_{p}q^{2}-(\sigma_{\pi \pi}/32 \pi^{2}) h_{1}(q^{2}).
   \label{AGa}
\end{eqnarray}
where
 \begin{eqnarray}
 &&   h_{1}(q^{2})  = \frac{q^{2}}{\pi} \\ \nonumber
 && [ \frac{8 \mu^2}{q^2} - (\frac{4 \mu^2+q^2}{q^2})^{3/2}
    ln\frac{\sqrt{4\mu^2+q^2} +q}{\sqrt{4\mu^2+q^2} -q}  +ln\frac{m^{2}}{\mu^{2}}  ] ,
   \label{Aga-h1}
\end{eqnarray}
with $q^2=-t$ . Note that we have  removed
the misprint in this equation, as made in \cite{Khoze-Sl00,Khoze-Sl14}.
They obtained the limits of the representation in the brackets at
$q^{2} >> 4\mu^{2} $; hence, the slope grows in order $q^{2}$  with small logarithmic suppression.
 At small $t$  ($q^{2} \ll 4\mu^{2}$ ) they predicted that the representation in the brackets goes to
  $ln(m^{2}/\mu^{2})-8/3$. 
  Note that the authors aimed to explain the deviation of the slope from the  constant
  at a non-small momentum transfer (in the domain $-t=0.4$ GeV$^2$). However, in this domain the
  impact of the diffraction minimum is already felt. Hence this domain of $t$ is usually described by
  an additional term like $c t^2$, which  were proposed earlier (for example \cite{VHove-lect}).


\begin{figure}
\begin{center}
\includegraphics[width=0.45\textwidth] {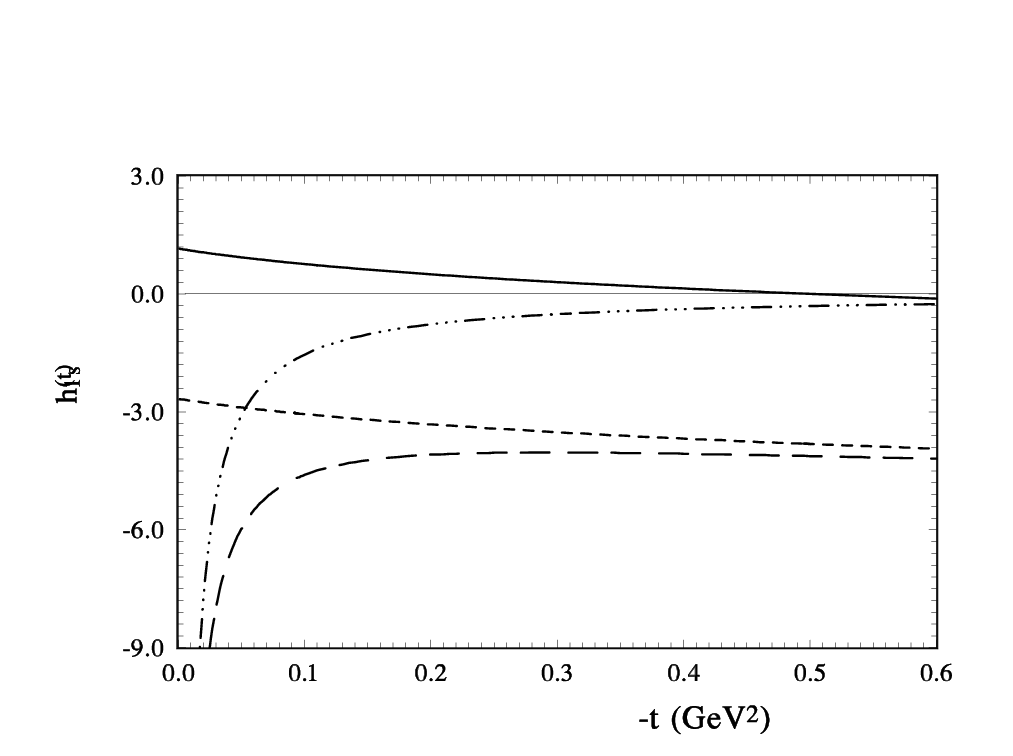}
\end{center}
\caption{
 The $t$ dependence of the different parts of the terms of eq.(9) 
 (solid line - the sum of all three terms in the brackets;
 the  dots-dashed line - the first term is multiplied by ($-1$);
 the long-dashed line - the second term, the short-dashed line  - the sum of the two first terms).
 }
\end{figure}

\vspace{0.5cm}

\begin{figure}
\begin{center}
\includegraphics[width=0.4\textwidth] {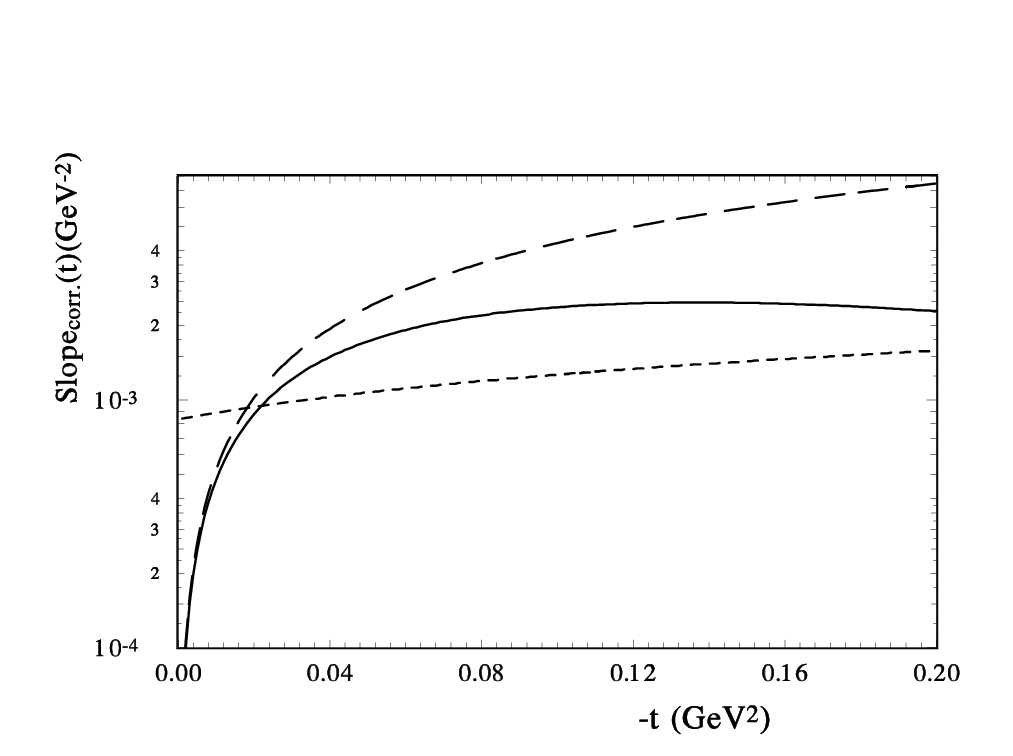}
\end{center}
\caption{
 The $t$ dependence of the additional terms to $\alpha^{\prime} t$
 (solid line - eq.(9);   
 long-dashed line - eq.(\ref{stsq});
 short-dashed line -  eq.(\ref{stsq}) without extraction of $2 \mu$  and multiplied by $0.1$.
 }
\end{figure}

  Practically at the same time a similar equation   was obtained in \cite{Barshay-72}
   \begin{eqnarray}
   D_{R}^{NN}(t) = n [A -
   (\frac{(4 \mu^2+q^2)^{3/2}}{q})
    ln\frac{\sqrt{4\mu^2+q^2} +q}{\sqrt{4\mu^2+q^2} -q}  ]^{-1} ,
   \label{Barshay-72}
\end{eqnarray}
 They proposed to approximate this representation by
  \begin{eqnarray}
  D_{R}^{NN}(t) \sim [\frac{1}{5 \mu^2 -t} + C_{const}]
  \label{Drt}
\end{eqnarray}
  For numerical calculation and comparison with experimental data, they took
  $C_{const}=24.3$ GeV$^{-2}$.

  In \cite{Khoze-Sl00}, using the calculations \cite{Gribov-Sl72} and removing the misprint,
  the authors  proposed for $h_{1}$,
  with taking into account the meson form factor (with $\Lambda_{\pi}^2=m_{\rho}^{2}$ GeV$^2$ ), the following equation:
   \begin{eqnarray}
   h_{1}(q^{2}) = &&\frac{4}{\tau} f^{2}_{\pi}(t) [2 \tau - (1+\tau)^{3/2} \\ \nonumber
    ln\frac{\sqrt{1+\tau} +1}{\sqrt{1+\tau} -1}
    && +ln\frac{m^{2}}{\mu^{2}}  ] ,
   \label{h1ff}
\end{eqnarray}
where $\tau = 4 \mu^2/q^{2}$ and
 \begin{eqnarray}
  f_{\pi}(t) =  \frac{\Lambda_{\pi}^2}{ \Lambda_{\pi}^2 -t}
  \label{ffpi}
\end{eqnarray}

   Really, for the slope, which is multiplied by $q^2$, there are two divergence terms with different signs
    plus a small constant term. The divergence terms cancel each other and the rest have a slow $t$ dependence (see Fig. 2).
   The   cancelation of the two terms as $t \rightarrow 0$ is not full and the rest
   have also some indefiniteness and strongly depend on $t$.
   In Fig. 2, the $t$ dependence of the different parts of the term of eq.(9) 
    in the brackets
     is shown. The cancelation of the two diverse terms leads to the negative contribution to the standard constant slope.
     However, the sum of all three terms gives a positive additional contribution with a small $t$ dependence.

   In  Fig. 3, we show the $t$ dependence of a different form of the slopes with the full kinematic coefficient.
  One can see  a very different $t$ dependence , especially in the form $\sqrt{4\mu^2+q^2} $ which was proposed
   in \cite{Jenk-72} without the extraction of $2 \mu$.
      Some models of  hadron interactions at high energy
       suppose that
       the slope has a slow $t$ dependence.
 For example, the  Dubna dynamical model (DDm) \cite{DDM},  which takes into account the contribution to  hadron
   interactions from the meson cloud of the nucleon and  uses the standard eikonal form of the unitarization,
    leads to the scattering amplitude in the form
  \begin{eqnarray}
 T(s,t) =&& -is \sum_{n=1}^{\infty} \frac{\mu}{(n^2 \mu^2 -t)^{3/2}} \\ \nonumber
  && (1- b\sqrt{n^{2}\mu^{2}-t})e^{-b\sqrt{n^{2}\mu^{2}-t}}.
\end{eqnarray}
   The analysis of the high energy data on the proton-antiproton scattering in the framework of this model
   shows the obvious non-exponential behavior of the differential cross sections. 
   For $\sqrt{s} = 540 $ GeV it shows the change in the slope
   from $16.8$ GeV$^{-2}$ at $t=-0.001$ GeV$^2$ up to $14.9$ GeV$^{-2}$ at $t=-0.12$ GeV$^2$. For the same $t$ the size of $\rho(s,t)$ changes
   from $0.141$ up to $ 0.089$.
   For Tevatron energy $\sqrt{s} = 1800 $ GeV it shows the change of the slope
   from $18.1$ GeV$^{-2}$ at $t=-0.001$ GeV$^2$ up to $15.9$ GeV$^{-2}$ at $t=-0.12$ GeV$^2$ and again  the size of $\rho(s,t)$ changes
   from $ 0.182$ up to $ 0.143$. Hence, the model shows the continuously decreasing  slope and $\rho$ at small $t$.
   Practically the same results were obtained in \cite{Pumplin-92} in the framework of the model that also uses the eikonal unitarization.

   The analysis \cite{Sel-UA42} of the high precision data obtained at $SP\bar{P}S$ at $\sqrt{s} =541$ GeV in the UA4/2 experiment
   shows the existence in the slope of the term proportional to $\sqrt{|t|}$. This  term can be related with the nearest
   $\pi$ meson threshold or, as was shown in \cite{Sel-UA42}, such behavior of the differential cross sections can reflect
   the presence of the contribution of the spin-flip amplitude.
  As was noted in  \cite{Rev-LHC}),
 the analytic $S-$matrix theory, perturbative quantum chromodynamics
and the data require  Regge trajectories to be nonlinear complex
functions \cite{nonlin1,Jenk-nl}. 
  The Pomeron trajectory has threshold singularities, the lowest one
being due to the two-pion exchange required by the $t-$channel
unitarity \cite{Gribov-Sl72}.
This threshold singularity appears in different forms in various
models (see \cite{Rev-LHC}).
  In the recent high energy general structure model (HEGS) \cite{HEGS1}, 
     a  small additional term is introduced into the slope which reflects some possible small nonlinear
        properties of the intercept.
     As a result, the slope is taken in the form
   \begin{eqnarray}
   B(s,t) \ =  -\alpha^{\prime} \ln(\hat{s})  [ 1 - d_1 t/\ln(\hat{s})  e^{ d_2 \alpha_{1}  t \ln(\hat{s})} ]  .
   \label{B0}
\end{eqnarray}
   This form leads to the standard form of the slope as $t \rightarrow 0$ and $t \rightarrow \infty$
   Note that our additional term at large energies has a similar form as the additional term to the slope
   coming from $\pi-loop$ examined in \cite{Gribov-Sl72} and recently in \cite{Khoze-Sl00}.

\section{Experimental database}

   In our research, we use the widest 
    region of experimental data.
   The energy region begins from $\sqrt{s} = 3.5 - 3.8$ GeV for proton-antiproton scattering.
   At these energies  new data were obtained in the high precision experiment on elastic $p\bar{p}$ scattering
   at small angles. It has four sets at different energies in the momentum transfer region
   $|t|= [0.000986 - 0.02]$ GeV \cite{pap3p6}. This experiment is very important as they obtained the value of $\rho(s,t=0)$
   with a remarkably small error and with the size near zero. It is essentially different from the
   value of  $\rho(s,t=0)$  obtained in the framework of the dispersion relation analysis by P. Kroll,
   which was carried out on the basis of old experimental data with large errors.

   Low-energy proton-antiproton data from $\sqrt{s} = 11.54$ GeV
   up to the final  ISR energy $\sqrt{s} = 62$ GeV are represented in eleven sets \cite{HEP-data}.
   Then we include seven sets of experimental  data obtained at the $SP\bar{P }S$ collider at energies around $\sqrt{s} = 540-630$ GeV.
   The  Tevatron data  at $\sqrt{s} = 1800 - 1960$ GeV are represented in four sets.
    The latter data present the maximal energy obtained for the proton-antiproton scattering at accelerators.
   On the whole, for the $p\bar{p}$ elastic scattering we have 33 sets of the different experiments.

    For the proton-proton elastic scattering we take into account  sixty five sets of different experiments
    from the low energy  $\sqrt{s} = 6.1$ GeV up to the maximal  LHC energy $\sqrt{s} = 13$ TeV  \cite{Spires-data,Land-Bron}.
    We especially note the high precision experimental data obtained at a small momentum transfer by the FNAL
    collaborations at $\sqrt{s} = 9.8, 9.9, 10.6,12.3 $ GeV and at $\sqrt{s} = 19.4, 22.2, 23.9, 27.4$ GeV
    which start from a very small momentum transfer $t=-0.00049$ GeV$^2$.
    They can be compared with
    the data obtained by the UA4/2 Collaboration at the $SP\bar{P}S$ collider, which start from
     $t=-0.000875$ GeV$^2$ and with the data obtained at the LHC by the TOTEM Collaboration
     at $\sqrt{s} = 13$ TeV which reached the  $t=-0.00029$ GeV$^2$.
      Of course, the last case achieved the smallest possible
     angles of scattering.  We take into account  the experimental data at a high value of momentum transfer
     up to momentum transfer $-t=10 - 14$ GeV$^2$. The corresponding experimental data were obtained at
     energies $\sqrt{s} = 19.4, 27.4, 52.8$ GeV.

     Summarizing all the sets of experimental data on  elastic scattering at not large angles,
     we took into account  $115$ sets of  different experiments which included $4326$ experimental points on proton-proton and proton-antiproton elastic scattering.

       We included  in our simultaneous analysis the data  for the spin correlation parameter $A_N(s,t)$ of the polarized proton-proton
       elastic scattering. This set of data includes 235 experimental data for relatively small energies
       $\sqrt{s}= 3.63$ GeV and up to  $\sqrt{s}= 23.4$ GeV. During our fitting procedure sufficiently good
       descriptions  of the experimental data were obtained.

      For the first time, we also included in our research the elastic proton-neutron experimental data.
     The corresponding data were taken into account beginning from the   $\sqrt{s} = 4.5$ GeV up to maximal
      energy proton-neutron collisions  $\sqrt{s} = 27.19$ GeV obtained at accelerators.
     It should be noted that such energy represented an average between minimum and maximum energies.
     For example, maximum energy obtained at the FNAL presented average between
     $P_L=340$ GeV/c and $P_L=400$ GeV/c.
     It is very important that in such experiments a very small
     momentum transfer was reached, for example $-t_{min} =  0.23 \ 10^{-4}$ GeV$^2$ at   $\sqrt{s} =  23.193$ GeV.
       On the whole,  we took into account $24$ sets of  experimental data of different experiments
      which supply $526$  experimental data. Hence, on the whole, we took into our analysis
       $5027$ experimental data on  elastic nucleon-nucleon scattering.

\section{Main amplitudes  of the high energy generalized structure (HEGS) model}

\subsection{Asymptotic  part of the scattering amplitude}
 The model is based on the idea that at high energies a hadron interaction in the non-perturbative regime
      is determined by the non-perturbative Regge gluons
      exchange. The cross-even part of this amplitude can have two non-perturbative parts, possible standard
      non-perturbative QCD pomeron - $(P_{2np})$ and cross-even part of  3-non-perturbative gluons ($P_{3np}$).
      The interaction of these two objects is proportional to two different form factors of the hadron and
      both form factors are discussed in the next subsection.
      This is the main assumption of the model.
      Both terms have the same intercept.
       This corresponds to the maximal 
       Odderon, which was introduced by
      L. Lukazsuk, B. Nicolescu \cite{Luk-Nic}.

 Note, the  Odderon
  that arises naturally in perturbative QCD, for example
  \cite{KovLevin}, with the intercept $\alpha_{Odd}(0) = 1$. 
 They show that shadowing corrections decreased Odderon contributions.
  However, the experimental data show the Odderon contribution at ISR energy $\sqrt{s}=52.8$ GeV and
   at Tevatron energy \cite{Royon} and probably at LHC energy $\sqrt{s}=13$ TeV. 
  For example, in \cite{Khoze 1801.07065} it was noted
"On the other hand, it is possible to introduce the Odderon phenomenologically as an object
which does not violate first principles and the axiomatic theorems. In fact it was stated in \cite{NM-rho}
that the new TOTEM result is a definitive confirmation of the experimental discovery of the
Odderon in its maximal form".

      In our fitting procedure 
       we checked up such a possibility and made the fit with two intercepts.
  As a result we obtained the Odderon intercept $\alpha_{0-Odd}-1=0.1101\pm 0.0004$
  (with fixed Pomeron intercept $\alpha_{0-Pom}-1=0.11$).
  and  that $\chi^2$  did not practically  changes.
  Hence, this  confirms our assumption about the equality of  both intercepts.

   The second important assumption is that we choose the slope of the second term
       four times smaller than the slope of the first term, by  analogy with the two pomeron cuts.
      An estimation of 
  the diffraction slope of the generalized BFKL pomeron  was made in \cite{NNN-Sl}
in terms of the correlation radius for the perturbative gluons.
They show that indeed the pomeron trajectory has the finite slope $\alpha^{\prime}_{IP} \sim R^{2}_{c}$ ,
where $R_{c}$ 
is the correlation radius for the perturbative gluons.
They note that in the gBFKL dynamics 
a dimensionful $\alpha^{\prime}_{IP}$  is a noneperturbative quantity
related to the nonperturbative infrared parameter of the model - the gluon propagation
radius $Rc$. For the preferred $Rc = 0.27$ fm, quite a small $\alpha^{\prime}_{IP} = 0.072$ GeV$^{?2}$ is found.
  In our work we used the phenomenological properties of the Pomeron (Odderon)
  taking into account that the Odderon only feels the
centre of the proton and not the pion cloud. Therefore, it is reasonable to assume that the Odderon slope,
 $B_{Odd}$, is lower than that for the even-signature (Pomeron) amplitude \cite{Khoze  1801.07065}.
    This takes into account the data of the ISR  experiments which show
that the slope of the differential cross sections after the second bump
is approximately four times smaller than the slope at small $t$
as well as the conclusion of P. Landshoff that the Odderon contributions
     are  essentially important at large momentum transfer.
       Our fitting procedure confirms also such assumption when we take the second slope as a free parameter.

\subsection{Form factors}

\begin{figure}
		\includegraphics[width=0.3\textwidth] {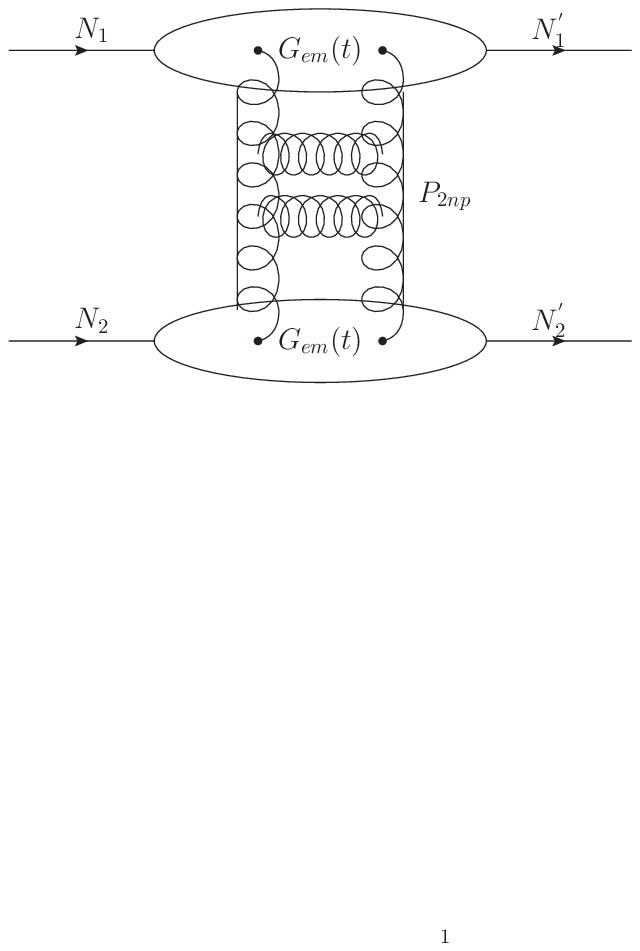}

\vspace{-4.cm}
		\includegraphics[width=0.3\textwidth] {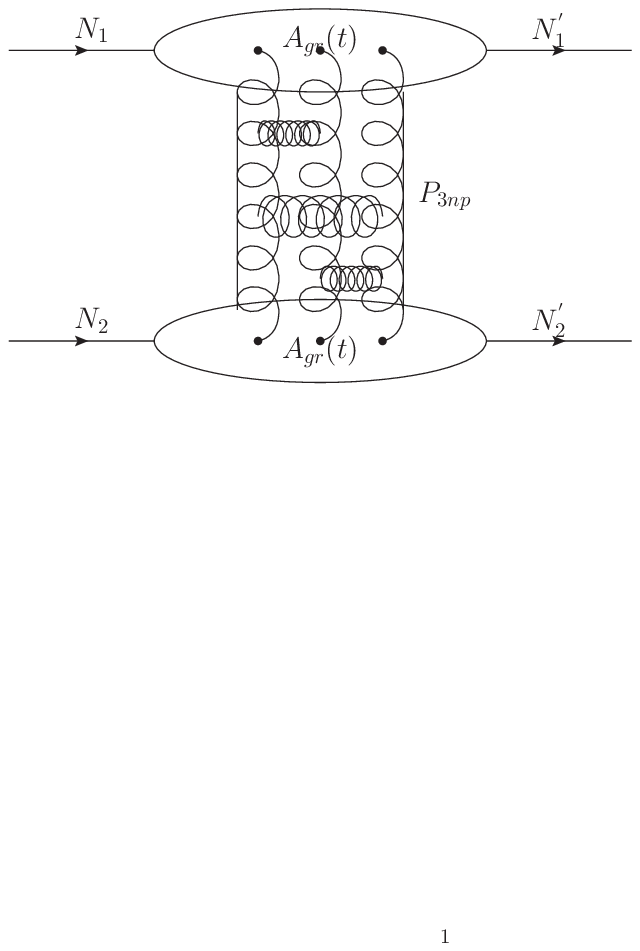}
\vspace{-4.cm}
	\caption{The Born amplitudes of nucleon-nucleon elastic scattering:
 (top) with two non-perturbative gluons - $P_{2np}$ and (bottom) three non-perturbative gluons - $P_{3np}$.
	}
	\label{fig:4}       
\end{figure}


     Since nucleons are not point particles,
   their structure must be taken into account.
    Usually, especially in the 60s - 70s, the researchers  proposed that
    the hadron form factor is proportional to the charge distribution into the hadron,
     which can be obtained from electron-nucleon scattering.

       This is primarily due to the electromagnetic structure of the nucleon
   which can be obtained from the electron-hadron elastic scattering.
   In the Born approximation,
       the Feynman amplitude for the elastic electron-proton scattering is  
  \begin{eqnarray}
M_{ep\rightarrow ep} = \frac{1}{q^2}[e \bar{u}(k_2)\gamma^{\mu} u(k_{1}][e\bar{U}(p_{2}\Gamma_{\mu}(p_{1},p_{2})U(p_{1}],
 \end{eqnarray}
   where $u$ and $U$ are the electron and nucleon  Dirac spinors,
\begin{eqnarray}
\Gamma^{\mu} = F_{1}(t)\gamma^{\mu} \ + \  F_{2}(t) \frac{i\sigma^{\mu\nu} q_{\nu}}{2 m},
\end{eqnarray}
   where $m$ is the nucleon mass, $\kappa$ is the anomalous part
   of the magnetic moment and $t=-q^2 =-(p-p^{\prime})^2$ is the square of the momentum transfer of the nucleon.
   The functions $F_{1}(t)$ and $F_{2}(t)$ are named  the Dirac and Pauli form factors, which depend upon the nucleon structure.

     However, it is not obvious that  strong interactions have to be proportional
     to the electromagnetic properties of  hadrons.
     Taking into account  this fact in  one of the famous models, Bourrely-Soffer-Wu \cite{soff}
     used some modification of the form factor with free parameters
     which was obtained from the description of the differential cross section
     of  hadron scattering. For example, it takes   the form
        \begin{eqnarray}
   G(t) =\frac{1}{1-t/m^{2}_{1}} \frac{1}{1-t/m^{2}_{2}} \frac{a^2+ t}{a^2- t}.
   \label{ffBs}
\end{eqnarray}
  However, this  allows some freedom in the $t$-dependence of the scattering amplitude.
  It is necessary to  take into account the parton distribution (PDFs) in the hadrons.
   However, PDFs  depend on the  Bjorken variable $x$.
   In the 80s some relations between PDFs and hadron form factor were proposed.

 Note that the function like GPDs(x,t, $\xi=0$) was used already in the old
    "Valon" model proposed by Sanevich and Valin in 1986 \cite{Valin}.
  In the model, the hadron elastic form factor was obtained by the integration function
  $L(x) G(x,t)$ where $L(x)$ corresponds to the parton function and $G(x,t  )$ corresponds to the
  additional function which  depends on momentum transfer and $x$.
  In modern  language,
  $L(x) G(x,t)$ exactly corresponds to the GPDs. The scattering amplitude is 
  \begin{eqnarray}
   M_{AB}(s,t) = K_A(q^2) K_B(q^2) V(s,q^2);
\end{eqnarray}
where $V(s,q^2)$ is a potential of strong interaction and $K_{AB}(q^2)$ are the corresponding form factors.
\begin{eqnarray}
   K_{p}(q^2) = \frac{1}{3} \int_{0}^{1}  dx [2 L_{p}^{u,d}(x) T_p^{u,d}(\vec{k});
\end{eqnarray}
where
$\vec{k} = (1-x) \vec{q}$, $k^2=(1-x)^2 q^2$ and
$$T_p^{u}(\vec{k}) = e^{6.1 k^2 }; T_p^{d}(\vec{k}) = e^{3 k^2 }$$.
 This form factor can be obtained taking into account PDF of   interacting  particles 
  which is multiplied by
 some function depending on momentum transfer $t$ and  Boirken variable $x$.

 Many different forms
  of the $t$-dependence of GPDs were proposed  \cite{Liuti2}.
   In the quark di-quark model  \cite{Liuti1}
    the form of  GPDs
   consists of three parts - PDFs, function distribution and Regge-like function.
 In other works (see e.g. \cite{Kroll04}),
  the description of the $t$-dependence of  GPDs  was developed
  in a  more complicated picture using the polynomial forms with respect to $x$.

Commonly, the form of $GPDs(x,\xi,t)$ is determined  through the
  exclusive deep inelastic processes  of type $\gamma^*p\rightarrow Vp,$ where $V$ stands for a  photon
   or a vector meson. However,
  such processes have a narrow region of momentum transfer and
  in most models the $t$-dependence of GPDs is taken in  factorization form
  with the Gaussian form of the $t$-dependence.
  Really, this form of $GPDs(x,\xi,t)$ can not be used
  to build a space structure of  hadrons,
   as for that one needs to integrate over $t$ in a maximally wide region.

  The conjunction between the momentum transfer and the impact parameter allows one to obtain
   a space parton distribution that has probability conditions  \cite{Burk-15}.
   The connections between the deep-inelastic scattering, from which we can obtain the $x$-dependence of
   parton distributions, and the elastic electron-nucleon scattering, where the form factors of the nucleons
   are obtained, can be derived by  using the
  sum rules  
  \cite{Mul94,Ji97a,Ji97b,R97}. 
  The form factors, which are obtained in different reactions, can be calculated
  as   the  Mellin moments of GPDs. 
 Using the electromagnetic (calculated as the zero Mellin moment of GPDs) and gravitomagnetic form factors
 (calculated as the first moment of GPDs) in the hadron scattering amplitude, one can obtain a quantitative
 description of  hadron elastic scattering in a wide region of energy and transfer momenta.


 The proton and neutron Dirac form factors are defined as
 \begin{eqnarray}
 F_1^p(t)=e_u F_1^u(t)+e_d F_1^d(t), \\ \nonumber
  F_1^n(t)=e_u F_1^d(t)+e_d F_1^u(t),
 \end{eqnarray}
  where $e_u=2/3$ and
 $e_d=-1/3$ are the relevant quark electric charges.
 As a result, the $t$-dependence of the $GPDs(x,\xi=0,t)$
can be determined from the analysis of the nucleon form factors
for  which experimental data exist in a wide region of momentum transfer.
 This is a unique situation as it unifies  elastic and inelastic processes.

In the limit $t\rightarrow 0$, the functions $H^q(x,t)$ are reduced to
 usual quark densities in the proton: $$ {\cal\
 H}^u(x,t=0)=u_v(x),\ \ \ {\cal H}^d(x,t=0)=d_v(x)$$ with the
 integrals $$\int_0^1 u_v(x)dx=2,\ \ \ \int_0^1 d_v(x)dx=1 $$
 normalized to the number of $u$ and $d$ valence quarks in the
 proton.
 The energy-momentum tensor $T_{\mu\nu}$
    \cite{Pagels,Ji97a,Ji97b,R04}  
    contains three gravitation form factors (GFF)
     $A^{Q,G}(t)$, $  B^{Q,G}(t)$, and $ C^{Q.G}(t)$.  We will scrutinize the first one which corresponds to
     the matter distribution in a nucleon.
      This form factor contains the quark and gluon contributions
        $$A^{Q,G}(t)=A_{q}^{Q,G}(t)+A_{g}^{Q,G}(t).$$

To obtain the true form of the proton and neutron  form factors,
 it is important to have the true form of the momentum transfer dependence of GPDs.
Let us 
 choose  the $t$-dependence of  GPDs  in a simple form
$ {\cal{H}}^{q} (x,t) \  = q(x) \   exp [  a_{+}  \
   f(x) \ t ],                                     $
  with $f(x)= (1-x)^{2}/x^{\beta}$
   \cite{ST-PRDGPD}.
The isotopic invariance can be used to relate the proton and neutron GPDs;
hence, we have the same parameters for the proton and neutron GPDs.


   The complex analysis of the corresponding description of the electromagnetic form factors of the proton and neutron
    by  different  PDF sets  (24 cases) was carried out in \cite{GPD-PRD14}. These
   PDFs include the  leading order (LO), next leading order (NLO) and next-next leading order (NNLO)
   determination of the parton distribution functions.
    They used  different forms of the $x$ dependence of  PDFs. 
    We slightly complicated the form of GPDs  in comparison with the equation used in     \cite{ST-PRDGPD},  
   but it is the simplest one as compared to other works 
\begin{eqnarray}
{\cal{H}}^{u} (x,t) \  = q(x)^{u}  \   e^{2 a_{H}   f(x)_{u}  \ t };  \\ \nonumber
{\cal{H_d}}^{d} (x,t) \  = q(x)^{d}  \   e^{2 a_{H} f_{d}(x)  \ t };  \nonumber
\end{eqnarray}
\begin{eqnarray}
{\cal{E}}^{u} (x,t) \  = q(x)^{u} (1-x)^{\gamma_{u}} \   e^{2 a_{E}  \  f(x)_{u}  \ t }; \\ \nonumber
{\cal{E_d}}^{d} (x,t) \  = q(x)^{d}  (1-x)^{\gamma_{d}} \   e^{2 a_{E} f_{d}(x) \ t },
\label{t-GPDs-E}
\end{eqnarray}
 where \\
 $ f_{u}(x)=\frac{(1-x)^{2+\epsilon_{u}}}{(x_{0}+x)^{m}}, f_{d}(x)=(1+\epsilon_{0}) (\frac{(1-x)^{1+\epsilon_{d}}}{(x_{0}+x)^{m}} ).$

 The hadron form factors will be obtained  by integration  over $x$ in the whole range of $x$ - $(0 - 1)$.
 Hence, the obtained  form  factors  depend on the $x$-dependence of the forms of PDF at the ends of the integration region.
 The  Collaborations determined the  PDF sets  from the inelastic processes only in  some region of $x$, which is only
 approximated to $x=0$ and $x=1$.
   Some  PDFs  have the polynomial form of $x$ with
     different power.  Others have the exponential dependence of $x$.
  As a result, the behavior of  PDFs, when $x \rightarrow 0$ or $x \rightarrow 1$,  can  impact  the
    form of the calculated form factors.

      In that work, 24 different PDF were analyzed.
    On the basis of our GPDs with, for example, the PDFs
    ABM12 \cite{ABM12},    we calculated the hadron form factors
     by the numerical integration
   and then
    by fitting these integral results by the standard dipole form with some additional parameters
$$   F_{1}(t)  = (4m_p - \mu t)/(4m_p -  t ) \  \tilde{G}_{d}(t),   $$
  with
  $$
  \tilde{G}_{d}(t) = 1/(1 + a_{1} q +q^{2}/a_{2}^2 + a_{3}^3 q^3)^2 $$
   which is slightly  different from
  the standard dipole form by two additional terms with small sizes of coefficients
  ($a_{1}=0.06$, GeV$^{-1}$,  $a^{2}_{2}=0.78$ GeV$^2$, and $a^{3}_{3}=0.08$ GeV$^{-3}$ ).
  The matter form factor 
\begin{eqnarray}
 A(t)= && \int^{1}_{0} x \ dx
 [ q_{u}(x)e^{2 \alpha_{H} f(x)_{u} / t  } \\	\nonumber
 && + q_{d}(x)e^{ 2 \alpha_{H} f_{d}(x)  / t}  ] 
\end{eqnarray}
 is fitted   by the simple dipole form  $  A(t)  =  \Lambda^4/(\Lambda^2 -t)^2 $
 with $ \Lambda^2 =1.6$ GeV$^2$.
 These form factors will be used in our model of  proton-proton and proton-antiproton elastic scattering
        and further in one of the vertices of pion-nucleon elastic scattering.

  To check the momentum dependence of the spin-dependent part of GPDs $ E_{u,d}(x,\xi=0,t) $,
   we can calculate the magnetic transition
  form factor, which is determined by the difference of $ E_{u}(x,\xi=0,t) $ and $ E_{d}(x,\xi=0,t) $.
 For the magnetic $N \rightarrow \Delta $ transition form factor $G^{*}_{M}(t)$,  in the large $N_{c}$  limit,
the relevant $GPD_{N\Delta}$ can be expressed in terms of the isovector GPD
    yielding the sum rules  \cite{R04}

The  experimental data  exist up to
   $-t =8 $ GeV$^2$ and our results  show a sufficiently good coincidence with experimental data.
   It is confirmed that the form of the momentum transfer dependence of  $E(x,\xi,t)$ determined in our model  is right.

  Now let us calculate the moments of the GPDs with inverse power of $x$.
  This  gives us the
     Compton form factors.
    The results of our calculations of the Compton form factors
          coincide well with the existing experimental data.
         $R_{V}(t)$ and $R_{T}(t)$ have a similar momentum transfer dependence
         but  differ essentially in size.
  On the contrary, the axial form factor $R_{A}$ has an essentially different $t$ dependence.

  A good description of the variable form factors and elastic scattering of  hadrons
 gives  large support to  our determination of the momentum transfer dependence of GPDs.
 Based on this determination of GPDs,
    one can calculate the gravitomagnetic radius of the nucleon using the integral representation of the form factor
  and make the numerical differentiation over $t$ as $t \rightarrow 0$. This method allows us
  to obtain a concrete form of the form factor by fitting the result of the integration of
   the GPDs over $x$.
   As a result, the gravitomagnetic radius is determined as
   \begin{eqnarray}
<{r_{A}}^2> =  -\frac{6}{A(0)} \frac{dA(t)}{dt}|_{t=0};
\label{Rp}
\end{eqnarray}

 We used the same procedure as for our calculations of the matter radius.
  As a  result, the Dirac radius is determined from the zero Mellin moment of GPDs
   \begin{eqnarray}
<{r_{D}}^2> =  -\frac{6}{F(0)} \frac{dF(t)}{dt}|_{t=0};
\label{Rp1}
\end{eqnarray}
    where $F(t) = \int_{0}^{1} ( e_{u} \ q_{u}(x) \ + \ e_{d} \  q_{d}(x)) \ e^{-\alpha \ t f(x)} dx$.

  One can obtain  gravitational form  factors of quarks, which are related to the second  moments of GPDs.
 For $\xi=0$, one has
 \begin{eqnarray}
\int^{1}_{0}dx \ x{\cal{H}}_q(x,t) = A_{q}(t); \,\ \int^{1}_{0}  dx \ x {\cal{E}}_q (x,t) = B_{q}(t).
\end{eqnarray}
%
The parameters of the phenomenological form
  of GPDs can be obtained from the analysis of the experimental data for the
  proton and neutron electromagnetic form factors simultaneously.
  Our determination  of the momentum transfer dependence of GPDs of hadrons
  allows us to obtain  good quantitative descriptions of
   different form factors, including
    the Compton, electromagnetic,
     transition, and gravitomagnetic form factor  simultaneously.

\section{Model approximation and the structure of the elastic scattering amplitude}

  The differential cross sections of nucleon-nucleon elastic scattering  can be written as a sum of different
  helicity  amplitudes:
 \begin{eqnarray}
   \frac{d \sigma}{dt} = \frac{2 \pi}{s^2} ( |\Phi_{1}|^2 + |\Phi_{2}|^2+
   |\Phi_{3}|^2 + |\Phi_{4}|^2+4 |\Phi_{5}|^2.
\end{eqnarray}
      and the spin correlation parameter $A_{N}(s,t)$ is
\begin{eqnarray}
 && A_{N} \frac{s^2}{4\pi}\frac{d \sigma}{dt}  = \nonumber \\
  &&
  - [ Im (\Phi_{1}(s,t) + \Phi_{2}(s,t)+ \Phi_{3}(s,t) - \Phi_{4})(s,t) \Phi^{*}_{5}(s,t)].
  \label{AN}
\end{eqnarray}


  The HEGS model \cite{HEGS0,HEGS1} takes into account all five spiral electromagnetic amplitudes.
   The electromagnetic amplitude can be calculated in the framework of QED.
   For the spin-flip amplitudes,
   with the  electromagnetic and hadronic interactions included, every amplitude $\Phi_{i}(s,t)$
  can be described as
\begin{eqnarray}
  \Phi_{i}(s,t) =
  \Phi^{em}_{i} \exp{(i \alpha \varphi (s,t))} + \Phi^{h}_{i}(s,t),
\end{eqnarray}
where
 $  \varphi(s,t) =  \varphi_{C}(t) - \varphi_{Ch}(s,t)$, and
   $ \varphi_{C}(t) $ are calculated in the second Born approximation
 in order  to allow the evaluation   of   the Coulomb-hadron interference term $\varphi_{Ch}(s,t)$.
   The  quantity $\varphi(s,t)$
 has been calculated at a large momentum transfer including
  the region of the diffraction minimum 
  \cite{selmp1,PRD-Sum}. 


In the standard perturbative picture the spin-flip amplitudes die with growing of energy.
However, there are  different non-perturbative approaches which show non-dying spin-flip amplitudes,
for example, \cite{KopZak,Forte,zpc,Dor} 
One of these  is by E. Kuraev's \cite{G-Kuraev1,G-Kuraev2}, which we used for our non-dying part of the spin-flip amplitude.
 We used also the second part of the spin-flip amplitude which  has  a Regge form (eq.50) and has the energy dependence of order $1/s$.

\subsection{Electromagnetic amplitudes and phase factor}

   The electromagnetic amplitude can be calculated in the framework of QED
   in the one-photon approximation 
  \begin{eqnarray}
  \Phi^{em}_1(t) = \alpha f_{1}^{2} \frac{s-2 m^2}{t},  
   \Phi^{em}_3(t) = \Phi^{em}_1(t), \nonumber \\
  \Phi^{em}_2(t) = \alpha  \frac{f_{2}^{2}(t)}{4 m^2},
  \Phi^{em}_{4}(t) = - \phi^{em}_{2}(t),  \nonumber \\
  \Phi^{em}_5(t) = \alpha \frac{s }{2m \sqrt{|t|}} f_{1}^{2}.
  \end{eqnarray}

 The Coulomb-hadron interference phase is calculated with the dipole electromagnetic form factor \cite{PRD-Sum}.
  For calculation of the Coulomb-hadron phase we take the energy dependence of the slope
  in the form  $b_{sl}=6+0.75 \ln(s)$. \\

  We take 
      $ \Lambda^2=0.71 $;  and
     $  \gamma=.577215665 $. \\

     $   \phi_{a}=q^2 (2 \Lambda^2+q^2)/\Lambda^4 \ln{(\Lambda^2+q^2)^2/(\Lambda^2 q^2)}; $  \\
     $   \phi_{b}=(\Lambda^2+q^2)^2/(\Lambda^4 (4 \Lambda^2+q^2)^2 \ q \sqrt{4 \Lambda^2+q^2}); $ \\
     $  (4 \Lambda^4 (\Lambda^2+7 q^2)+q^4 (10 \Lambda^2+q^2)) \ln(4 \Lambda^2/(\sqrt(4 \Lambda^2+q^2) +q)^2) ;$ \\
     $   \phi_{c}=(2 \Lambda^4-17 \Lambda^2 q^2-q^4)/(4 \Lambda^2+q^2)^2 ;$ \\

     $   \phi_{1-3} =\phi_{a}+\phi_{b}+\phi_{c}$ ;   \\
   $  \phi_{CN}=-\ln{b_{sl} q^2/2.}+\gamma + \ln(1+8/(b_{sl} \Lambda^2)) - \phi_{1-3} )$  

  \subsection{The hadron scattering amplitude}

 Let us define the hadron spin-non-flip amplitude as
\begin{eqnarray}
  F^{h}_{\rm nf}(s,t)
   &=& \left[\Phi_{1h}(s,t) + \Phi_{3h}(s,t)\right]/2; \label{non-flip}
 \end{eqnarray}
  At small momentum transfer 
  in the CNI region, there are  two contributions coming
 from the electromagnetic and the strong interactions.

   On the one hand, the interference of such contribution  gives the possibility to determine the size of the real part of the scattering amplitude. On the other hand, to determine the form of the imaginary part of the hadronic amplitude,
 it is necessary to extract the Coulomb contribution and the interference term.
    As the electromagnetic amplitude and its  main contribution as 
     $t \rightarrow 0$   are well known,
     the measure of experimental data at very small $t$ gives the possibility to improve the normalization of the  differential cross sections.

 The existence of the different sets of  experimental data
 also allows one to improve separate normalization of experimental data.
  To compare different sets, some independent basis is needed.

  Let us take the calculations of the differential cross sections carried out in the framework of the new high energy
 generalized structure (HEGS) model  \cite{HEGS0,HEGS1} as such a  basis. 
 The model has only  a few free parameters and it quantitatively describes  experimental data
 in a wide domain of the momentum transfer, including the data in the CNI region,
 in a very wide energy region (from $\sqrt(s)=6 $ GeV up to LHC energies) simultaneously with the same numbering of the free parameters.
The  HEGS model assumes  Born terms for the scattering amplitude which get unitarization via
the standard eikonal representation to obtain   the full 
elastic scattering amplitude.

The scattering amplitude has exact $s\leftrightarrow u$ crossing symmetry as it is written
  in terms of the complexified Mandelstam variable $\hat{s} = s e^{-i\pi/2}$  that determines its real part.
The scattering amplitude also satisfies  the integral dispersion relation at large $s$.
It can be thought of as the simplest unified analytic function of its kinematic variables connecting different reaction channels
without additional terms for separate regions of momentum transfer or energy.
 Note that the model reproduces the diffraction minimum of the differential cross section
 in a wide energy region \cite{HEGS-min}.
  The HEGS model describes the experimental data at low momentum transfer, including the Coulomb-hadron interference region,
and hence 
 includes all five electromagnetic spin amplitudes and the Coulomb-hadron interference phase.

Let us determine the Born terms of the elastic nucleon-nucleon scattering amplitude 
    using  both (electromagnetic and gravitomagnetic) form factors
  \begin{eqnarray}
  F_{Pom2}^{Born}(\hat{s},t)=&& h_{Pom2}  \ G^2_{em}(t)  F_{a}(\hat{s},t) ; \\ \nonumber
 F_{Pom3}^{Born}(\hat{s},t)=&& h_{Pom3}  \ A^2_{gr}(t)  F_{b}(\hat{s},t); \\ \nonumber
  F_{Odd3}^{Born}(\hat{s},t)=&& h_{Odd3} \ A^2_{gr}(t)  F_{b}(\hat{s},t);
    \label{FB}
\end{eqnarray}
  where $F_{a}(\hat{s},t)$ and $F_{b}(\hat{s},t)$  have the standard Regge form: 
  \begin{eqnarray}
  F_{a}(s,t) \ = \hat{s}^{\alpha_{0}+\alpha_{1} t} ; 
  F_{b,c}(s,t) \ = \hat{s}^{\alpha_{0}+\alpha_{2} t}, 
 \end{eqnarray}
  with $   \hat{s}=s \ e^{-i \pi/2}/s_{0}$ ;  $s_{0}= 4 m^{2}_{p} \ {\rm GeV^2}$.
  The intercept of all main parts of the scattering amplitudes $1+\alpha_{0} =1.11$ was chosen
  as arithmetic means from different works including those on
      inelastic scattering. 
 Hence, at the asymptotic energy we have the universality
  of the energy behavior of the elastic hadron scattering amplitudes.
 The main part of the slope of the scattering amplitude has the standard logarithmic dependence on the energy
 $   B(s) = \alpha_{1,2} \ ln(\hat{s}) $
  with $\alpha_{1}=0.24$ GeV$^{-2}$  and  $\alpha_{2}=\alpha_{1}/4$ GeV$^{-2}$.
   It is taken with some correction (eq. 15).

     Both the hadron  electromagnetic and gravitomagnetic form factors were used
        in the framework of the high energy generalized structure   (HEGS)
        model  of  elastic nucleon-nucleon scattering.
        This allowed us to build the model with a minimum number of fitting
        parameters \cite{HEGS0,HEGS1,NP-HP}.
   The Born term of the elastic hadron amplitude can now be written as
  \begin{eqnarray}
&& F_{h}^{Born}(s,t)=h_1 \ G^{2}(t) \ F_{a}(s,t) \ (1+r_1/\hat{s}^{0.5})  \\ \nonumber
    && +  h_{2} \  A^{2}(t) \ F_{b}(s,t) \     \\ \nonumber
     && \pm h_{odd} \  A^{2}(t)F_{b}(s,t)\ (h_{as}+r_2/\hat{s}^{0.75}) (-t)/(1- r_d t) ), 
    \label{FB1}
\end{eqnarray}
 where both (electromagnetic and gravitomagnetic) form factors are used.
 The 
 corresponding parameters are determined by the fitting procedure. 

\begin{figure}[h]
%
\includegraphics[width=0.45\textwidth]{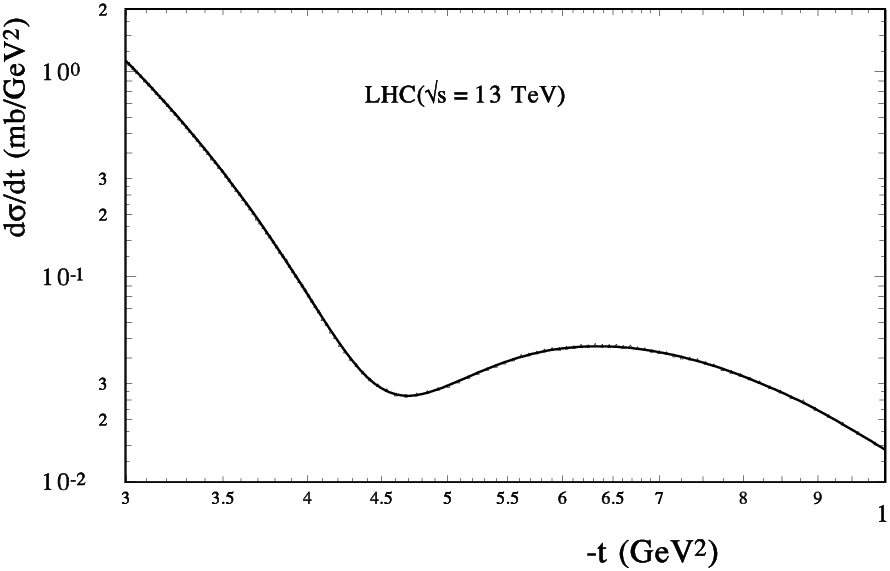}
\vspace{1.5cm}
\caption{ The  HEGS model calculation of $d\sigma/dt$ of  $pp $ scattering at $\sqrt{s} =13$ TeV
  (points  - experimental data of the TOTEM \cite{T67} Collaborations).
 }
\end{figure}

\begin{figure*}
%
 \includegraphics[width=0.9\textwidth]{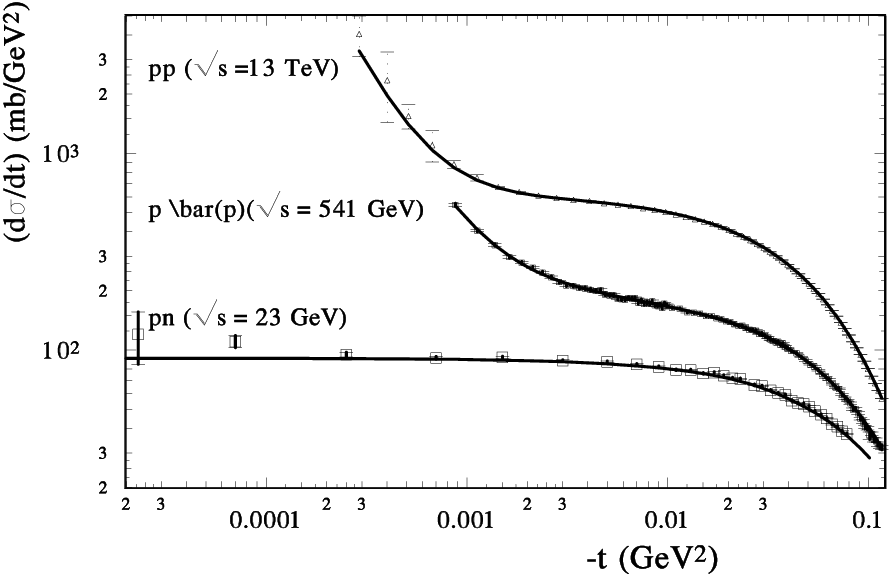}
\vspace{2.5cm}
\caption{ The comparison of the HEGS model calculation of $d\sigma/dt$ of  $pp $ scattering at $\sqrt{s} =13$ TeV
 (experimental data \cite{ATL13s},
 $p\bar{p}$ scattering at $\sqrt{s} =53$ GeV  (experimental data \cite{HEP-data}) and $pn $ scattering at $\sqrt{s} =23$ GeV
 (experimental data \cite{HEP-data}).
 }
\end{figure*}

The final elastic  hadron scattering amplitude is obtained after unitarization of the  Born term  \cite{cudpr-epja}.
There are some different approaches for summing the ladder diagrams
 - K-marix, U-matrix, standard eikonal, eikonal ( with some inelastic corrections \cite{Kaidalov}). 
 We analysed these forms and compared the results given by these approaches in some our works
 (for example, \cite{Sel-PL-B662,cudpr-epja,PRD79}
  In most part, at high energy ( especially at LHC energy) other approaches give  worse results.
  Moreover, we introduced  some parameters reflecting the Kaidalov corrections and obtained that
  their sizes are near zero or unity. So, this is conserves the standard eikonal   form.
  In a recent work   \cite{Luna-Rys} 
  the authors compare the U-matrix with the eikonal and write in the conclusion
  "As it is seen in Fig.3 the eikonal approach better agrees  with the ATLAS 13 TeV data."

    So, we use the eikonal representation and at first, we have to calculate the eikonal phase
 $$ \chi(s,b) \   =  -\frac{1}{2 \pi}
   \ \int \ d^2 q \ e^{i \vec{b} \cdot \vec{q} } \  F^{\rm Born}_{h}(s,q^2)  $$
  and then to obtain the final hadron scattering amplitude
 $$ F_{h}(s,t) = i s
   \ \int \ b \ J_{0}(b q) (1- \exp[ \chi(s,b)])    d b.$$
 The essential property of the model is that
 the real part of the scattering amplitude
 is obtained automatically through the complex $\hat{s}$ only.
  The scattering amplitude has exact $s\leftrightarrow u$ crossing symmetry.

\begin{figure}
\begin{center}
  \includegraphics[width=0.45\textwidth]{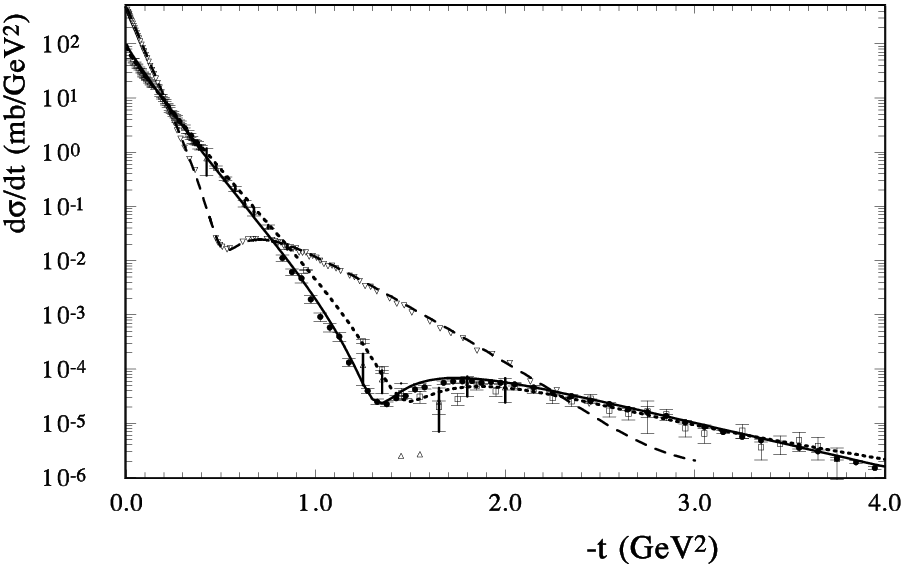}
\end{center}
\caption{ The comparison of the HEGS model calculation of $d\sigma/dt$ of  $pp $ scattering in the dip-bump regions
  at $\sqrt{s} = 7$ TeV (dashed line and open triangles down),
  $\sqrt{s} =53.8$ GeV (solid line and circles) and
  at $\sqrt{s} =19.4$ GeV ( tiny dashed line and boxes).
 }
\end{figure}

\begin{figure*}
\includegraphics[width=0.4\textwidth] {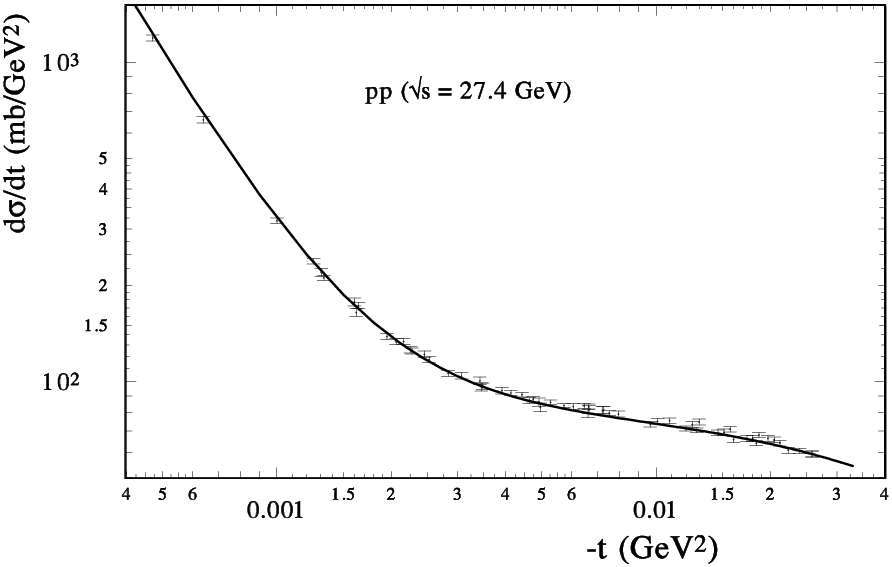}
\includegraphics[width=0.4\textwidth] {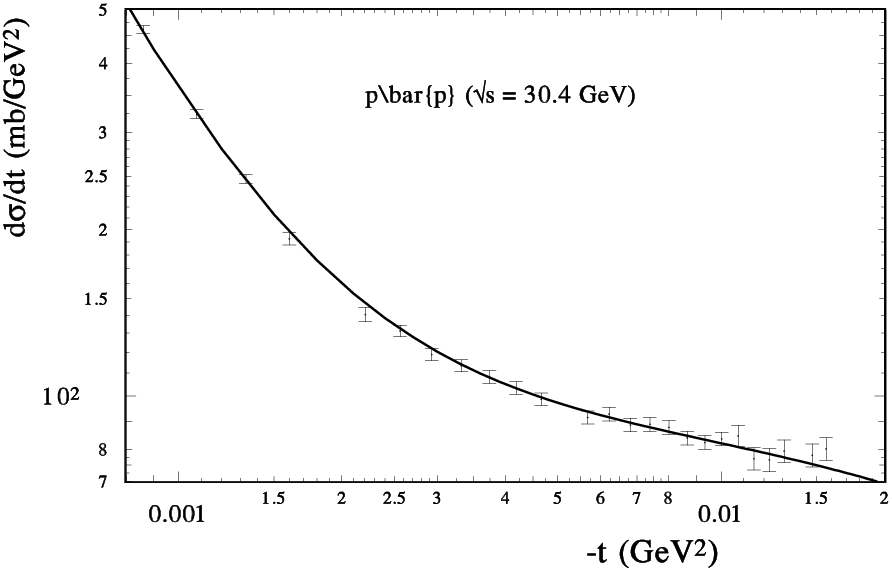}
\caption{  The comparison of the HEGS model calculation of $d\sigma/dt$ of  $pp $ scattering at $\sqrt{s} =27.4$ GeV [left]; \\
and  $p\bar{p}$ scattering at $\sqrt{s} =30$ GeV [right]. 
   }
\label{fig:7}       
\end{figure*}

   So to extend the model to low energies, it is necessary to take into account the contributions
   of the second Reggions. To avoid a substantially  increasing  number of fitting parameters,
   we introduce the effective terms which  represent the contributions of the sums
   of different Reggions.
    For the terms with the energy dependence order $1/\sqrt{\hat{s}}$
    we take for the $pp$ scattering the cross even part in the form
    \begin{eqnarray}
    F_{R1} = i h_{R1}/\sqrt{\hat{s}} e^{ b_{R1} t \ln{\hat{s}}} ;
    \end{eqnarray}
    and for  the $p\bar{p}$ the cross odd term in the form
    \begin{eqnarray}
    F_{R1} =  h_{R1}/\sqrt{\hat{s}} e^{ b_{R1} t \ln{\hat{s}}}
    \end{eqnarray}
   where the value $b_{R1}$ is fixed by unity.
   For the cross even part fast decreasing with growing energy,  we take a  term in the form
   \begin{eqnarray}
   F_{R2} = h_{R2}/\hat{s} e^{ b_{R2} t \ln{\hat{s}}} ;
   \end{eqnarray}
   As a result, the low energy terms require only three additional fitting parameters.

  The model is very simple from the viewpoint of the number of terms of the scattering amplitude   and    fitting parameters. 
  There are no any artificial functions or any cuts which bound the separate
  parts of the amplitude by some region of momentum transfer.
        In the framework of the model, the description of  experimental data was obtained simultaneously
        at the large momentum transfer and in the Coulomb-hadron region in the energy region from $\sqrt{s}=6 $ GeV
        up to LHC energies.  The model gives a very good quantitative description of the recent
        experimental data at $\sqrt{s}=13$ TeV in the region of the diffraction minimum \cite{Osc13} (see Fig. 5).

  As a result,  good descriptions of the experimental data of $pp$ and $p\bar{p}$ elastic scattering (90 experimental sets including 4326
  points and 256 polarization data) were obtained.
   Moreover a good description of $pn$ data was also  obtained.
  Some examples of such descriptions are presented  in Fig. 5-7.
  Of particular interest is Fig. 6 where
   the comparison of the HEGS model calculation of $d\sigma/dt$ of  $pp $ scattering at $\sqrt{s} =13$ TeV, $p\bar{p}$ scattering at $\sqrt{s} =53$ GeV, and $pn $ scattering at $\sqrt{s} =23$ GeV at small momentum transfer is shown.
    Practically for the first time, a  simultaneous research of
    proton-proton,  proton-antiproton and proton-neutron elastic scattering
    has been carried out in a wide energy (from $3.6$ GeV up to $13$ TeV)
    and momentum transfer region (from $|t| = 2. 10^{-4}$ GeV$^2$ up to  $|t| = 14$ GeV$^2$).

      In the fitting procedure based on  the modern version of FUMILY \cite{Sitnik1,Sitnik2},  we used only statistical errors.
  Systematic errors, which are mostly determined  by indefiniteness of  luminosity,
   were taken into account as an additional normalization coefficient,
   which is the same
  for all the data of a given set. The different normalization coefficients
  have practically random distributions.
   As a result, a wide range of possible forms of the scattering amplitudes
  prettily decreases.

\begin{figure*}
\includegraphics[width=0.4\textwidth] {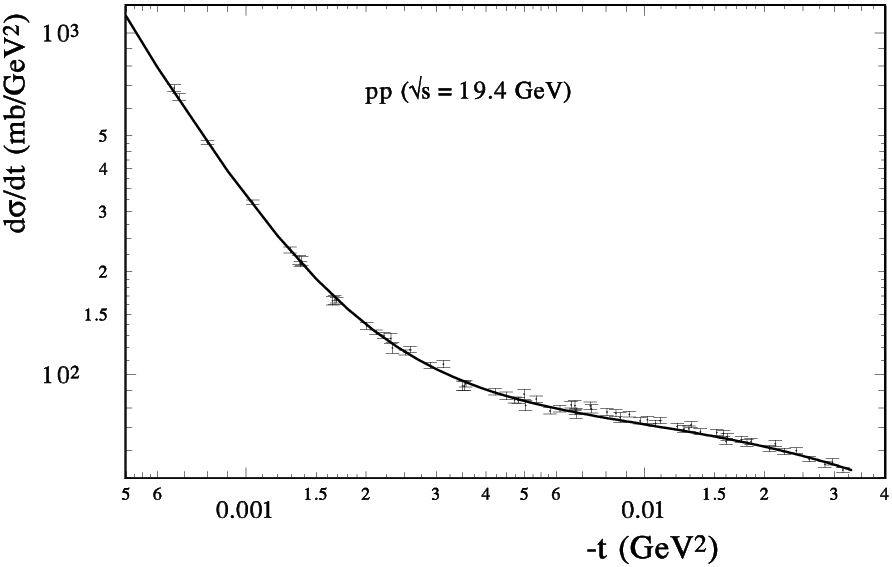}
\includegraphics[width=0.4\textwidth] {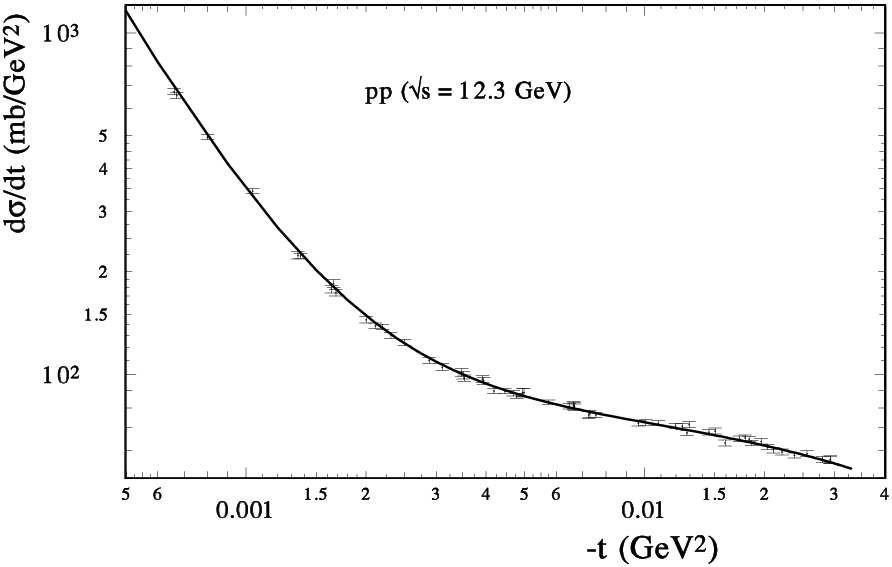}
\caption{The comparison of the HEGS model calculation of $d\sigma/dt$ of  $pp $ scattering at $\sqrt{s} =19.45$ GeV [left]; \\
and  at $\sqrt{s} =12.3$ GeV [right]. 
   }
\label{fig:17a}       
\end{figure*}

\begin{figure*}
\includegraphics[width=0.4\textwidth] {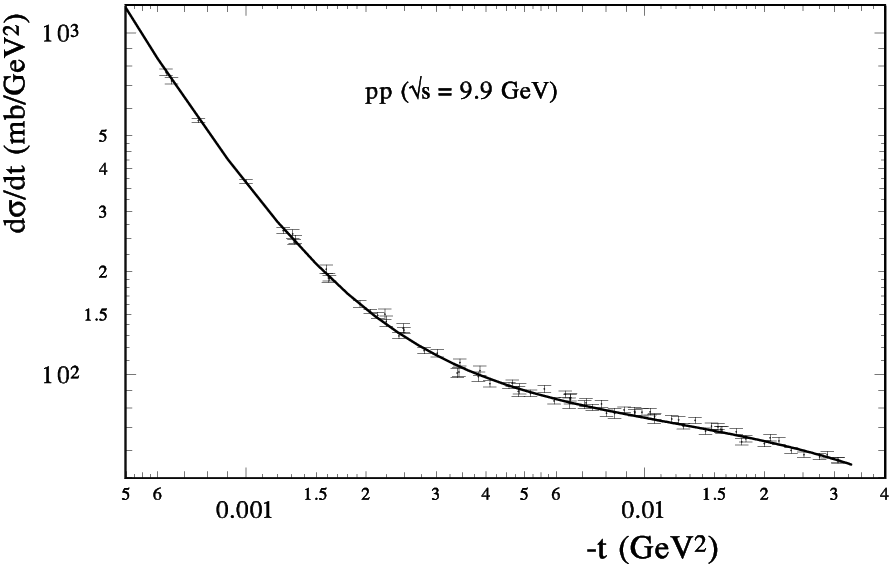}
\includegraphics[width=0.4\textwidth] {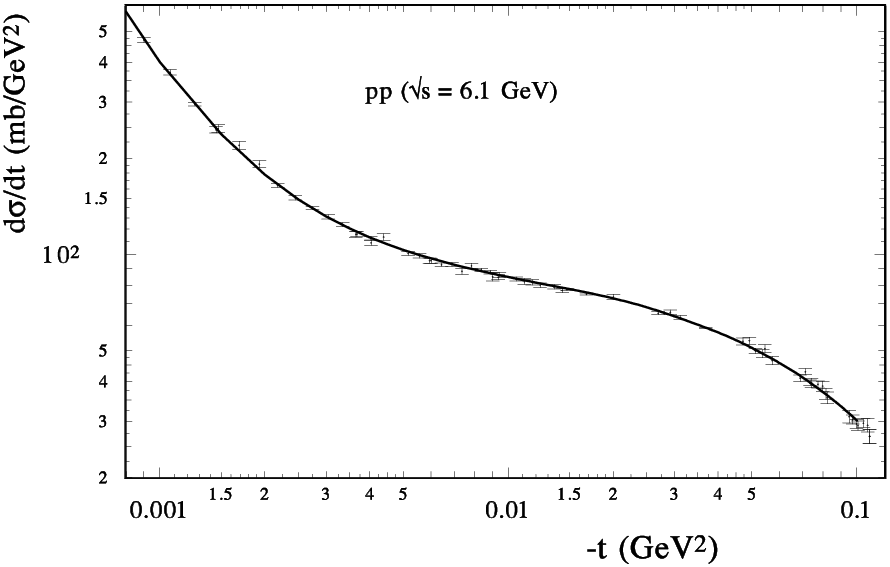}
\caption{The comparison of the HEGS model calculation of $d\sigma/dt$ of  $pp $ scattering at $\sqrt{s} =9.9$ GeV [left]; \\
and  at $\sqrt{s} =6.1$ GeV [right]. 
   }
\label{fig:9}       
\end{figure*}

\begin{figure}[t]
\includegraphics[width=65mm]{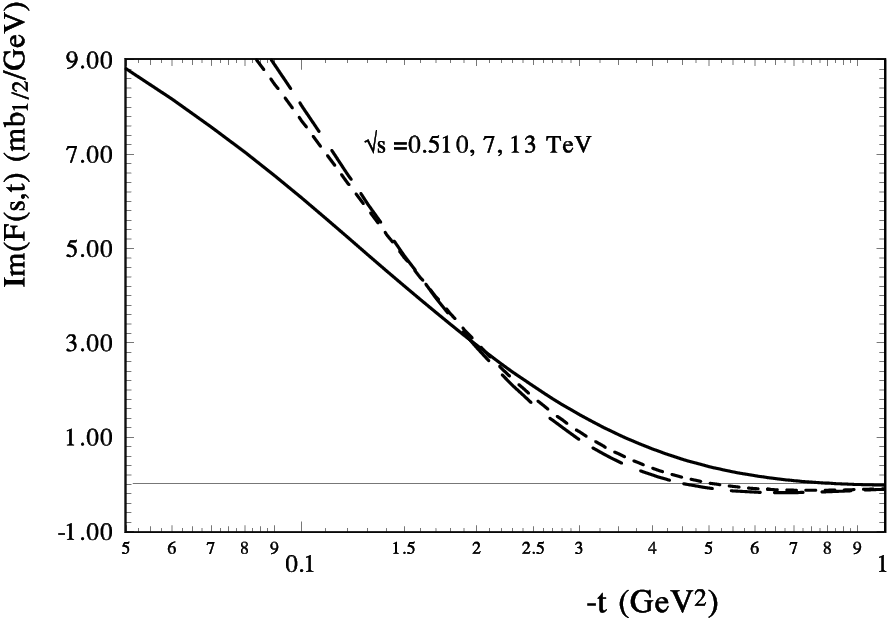}
\caption{The dependence of the imaginary part of the hadron scattering amplitude on $s$ and $t$
  calculated in the model at the energy  $\sqrt{s} =  0.51, 7, 13$ TeV }
\end{figure}

   A simultaneous  description of the  cross sections and spin correlation
   parameter of different nucleon-nucleon reactions, including 90 sets of experimental data,
   with the total number of data $N=4326$ gives a very reasonable $\sum_{i,j} \chi^2_{i,j} =4826$.
  The $pn$ case with $526$ experimental data, where the basic parameters were fixed from $pp$ and $p\bar{p}$ scattering,
   $\sum_{i,j} \chi^2_{i,j}=585$.


\begin{figure}
%
\includegraphics[width=0.4\textwidth]{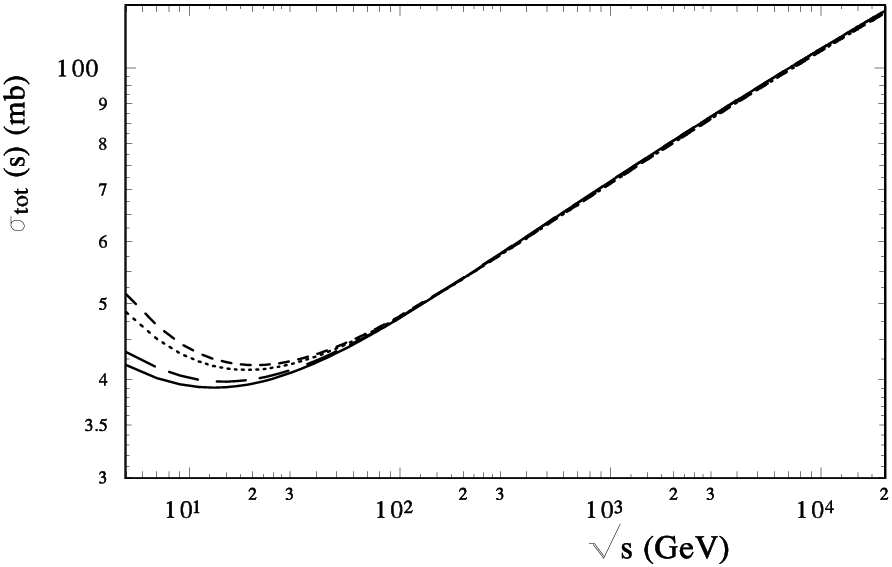}
\caption{  $\sigma_{tot}(s)$: dashed line - $p\bar{p}$ scattering, solid line -$pp$ scattering;
 the same but without the second Rejions contributions (tiny dashed - $p\bar{p}$ and long dashed $pp$ scattering).
  }
\label{Fig_11}
\end{figure}

\begin{figure}
\includegraphics[width=0.4\textwidth] {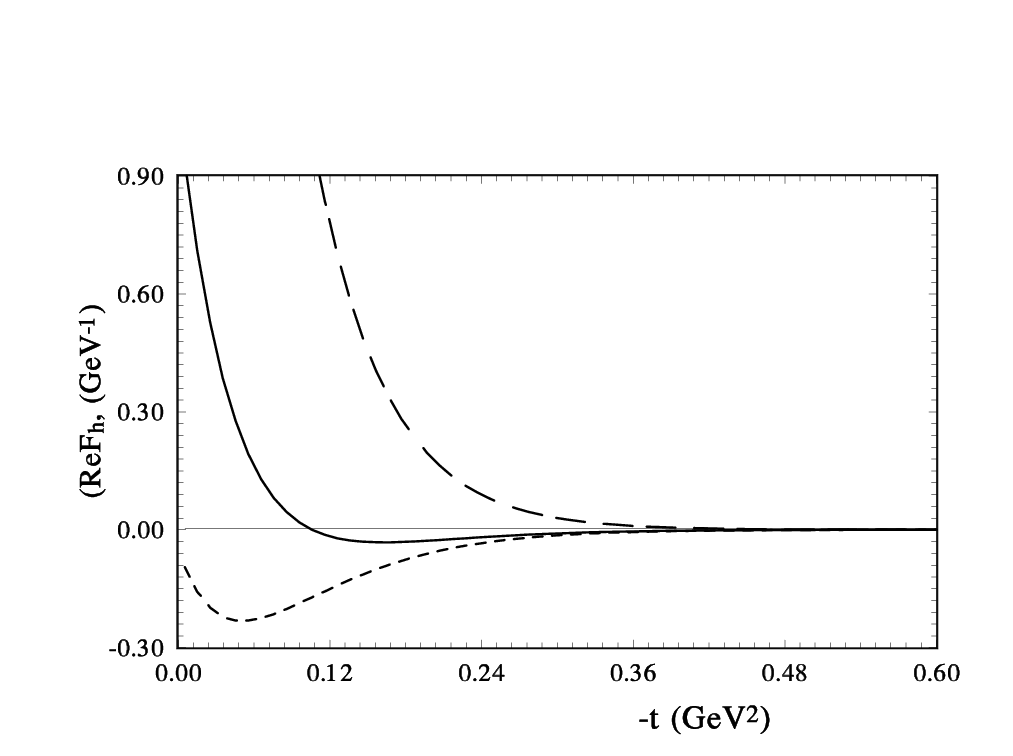}
\caption{ The real part of the scattering amplitude coming from complex $\hat{s}$ in the Regge representation
 (short dashed line - complex  $\hat{s}$ only in the exponential eq.(\ref{cExp});
  solid line full representation eq.(\ref{sscExp})
 }
\end{figure}

\begin{figure}[t]
\includegraphics[width=65mm]{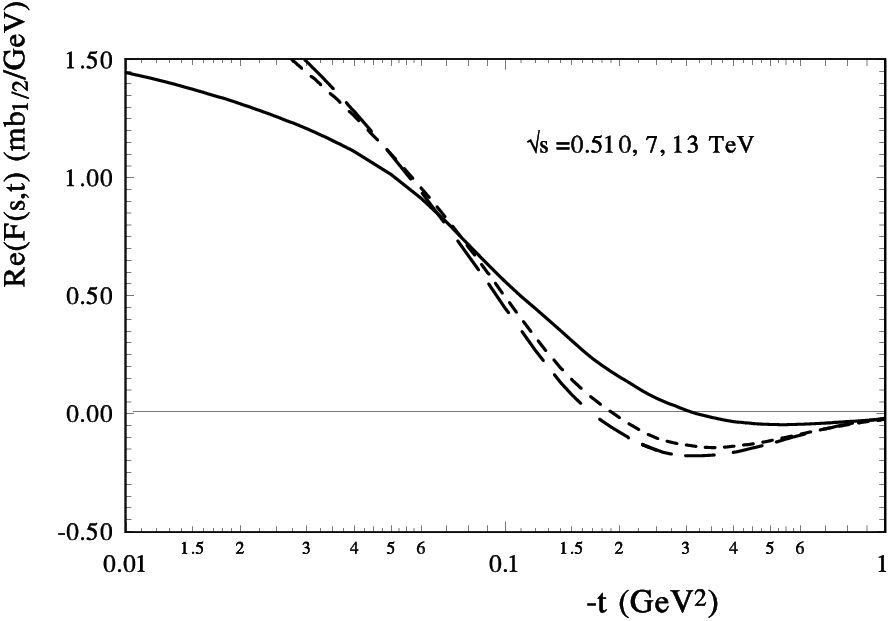}
\caption{The dependence of the real part of the hadron scattering amplitude on $s$ and $t$
  calculated in the model  at the energy  $\sqrt{s} =  0.51, 7, 13$ TeV }
\label{fig01}
\end{figure}

The data of the TOTEM at $\sqrt{s}=7 $ TeV are consistent and their mean value is equal to $98.5$ mb.
The ATLAS Collaboration, using their  differential cross section data in a region of $t$ where the Coulomb-hadron interference is
negligible, obtained the value $\sigma_{tot}=95.35 \pm 2.0$ mb. The difference between the two results, $\sigma_{tot}(s)$(T.) -
$\sigma_{tot}(s)$(A.) = $3.15$ mb, is about  1 $\sigma$.
 At $\sqrt{s}=8$ TeV, the measured value of $\sigma_{tot}$ grows,
  especially in the case of the TOTEM  Collaboration, 
  and the difference between the results of the two collaborations grows to
$\Delta(\sigma_{tot}(s)$(T.) - $\sigma_{tot}(s)$(A.) $= 5.6$ mb, i.e. 1.9 $\sigma$. This is reminiscent of
 the old situation with the measurement of the
 total cross sections at the Tevatron at $\sqrt{s} = 1.8$~TeV via the luminosity-independent method
by different collaborations.

Of course, we cannot say that the normalization of the ATLAS data is better
than that of the TOTEM data simply because it coincides with the
HEGS predictions.
 But this exercise may point to the main reason for the different values of the total cross sections obtained by  the
two collaborations.
This does not exclude some further problems with the analysis
of experimental data, e.g. those related to the analysis of the TOTEM data at $\sqrt{s}=7$ TeV \cite{Sel-NPA14}.

  One of the most important properties of the diffraction scattering is the form and energy dependence of the
  dip-bump structure. In Fig. 7, our model calculations are compared with the experimental data of $pp$ scattering at ISR energies
   ( $\sqrt{s}=19.4$ GeV and $\sqrt{s}=52.8$ GeV) and at the LHC energy $\sqrt{s}=7$ TeV and $\sqrt{s}=13$ TeV.
    At $\sqrt{s}=19.4$ GeV the real part of the $pp$ scattering changes its sign at $t=0$
    and we can see that it  gives a small contribution to the diffraction minimum too.
    It is interesting that at  $\sqrt{s}=13$ TeV the diffraction minimum
    has  a sharp form either. The ratio of the maximum to minimum of the differential cross section at $\sqrt{s}=19.4$ GeV
    is $2.4$
    and equals $1.8$ at the LHC energy.
    The position of the diffraction minimum $t_{min}(s,t)$ moves to low momentum transfer continuously  \cite{HEGS-min}
    and is described by the model very well (see, for example, Fig. 7 and Figs. 8,9,10 at low energies).
   It is interesting that the velocity of changing the position of the diffraction minimum
   changes very slowly. For example, from ISR energy $\sqrt{s}=53 $ GeV
   up to SPS energy $\sqrt{s}=540 $ GeV this position changes with a speed of $0.11$ GeV$^2$
    per $100$ GeV. Between $\sqrt{s}=540 $ GeV and $\sqrt{s}= 7 $ TeV
    such speed  is two times less and equals $0.006 $, at last  the position of minimum  between $7$ and $13$ TeV
     changes with a speed
    of  $0.002$ per $100$ GeV.
    The  scaling of this process can  approximately be represented as
       $t_{min} \ln{s/s_{0} } = const$.
  After the second bump the slope of the differential cross sections increases with energy.
    It  corresponds to the grows of the slope of the diffraction peak.
    In Figs. 7-9 the model calculations are compared with experimental data at low energies at small momentum transfer
    for $pp$ and $p\bar{p}$ scattering.
    It can be seen that the model reproduce well the existing experimental data
    of both reactions in that region.


   The point of $t$, where the imaginary part changes its sign, determines the position of the diffraction minimum.
   But it moves slightly at some large $t$ by   the contribution of the real part of the elastic hadron scattering amplitude.
 The behavior of the imaginary part of the scattering amplitude over momentum transfer calculated in the framework of the HEGS model
    is presented in Fig. 11 for the energies $\sqrt{s} =  0.51, 7, 13$ TeV.
    Again, we can see the point of crossover in the region of $|t|=0.2$ GeV$^2$.
    Despite the essential growth of the size of the imaginary part of the scattering amplitude at a very small momentum transfer, its slope slightly  changes with $t$ in the region of the  Coulomb nuclear interference.
     The size of the slope is practically proportional to the size of the total cross section in that region. However at larger $t$, for example at $|t|=0.1$ GeV$^2$,
    it grows essentially faster.

    It should be noted that the size of the slope of the differential cross sections  is determined in that region of
     momentum transfer by the CNI interference term which is proportional
    to $\alpha_{em}/t$ ( where $\alpha_{em}$ is a fine electromagnetic structure constant). It allows us to analyze  \cite{HEGS-AT13} the first points of the unique experiment carried out by the ATLAS Collaboration \cite{ATLAS-13}.

  The $s$ dependence of the total cross sections of $pp$  (solid line) and $p\bar{p}$ (dashed line) scattering is shown in Fig. 12.
   The difference of the $\sigma^{p\bar{p}}_{tot}$ and $\sigma^{pp}_{tot}$  is less than $1$ mb at $\sqrt{s}=20$ TeV.

\begin{figure}
\includegraphics[width=0.4\textwidth] {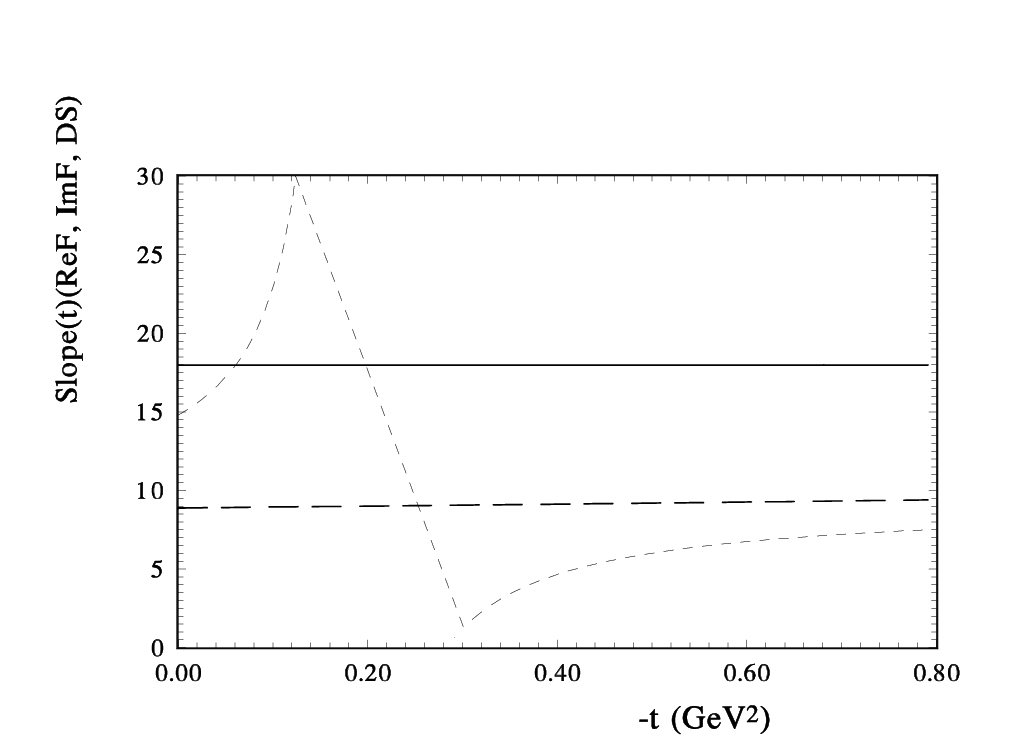}
\includegraphics[width=0.4\textwidth] {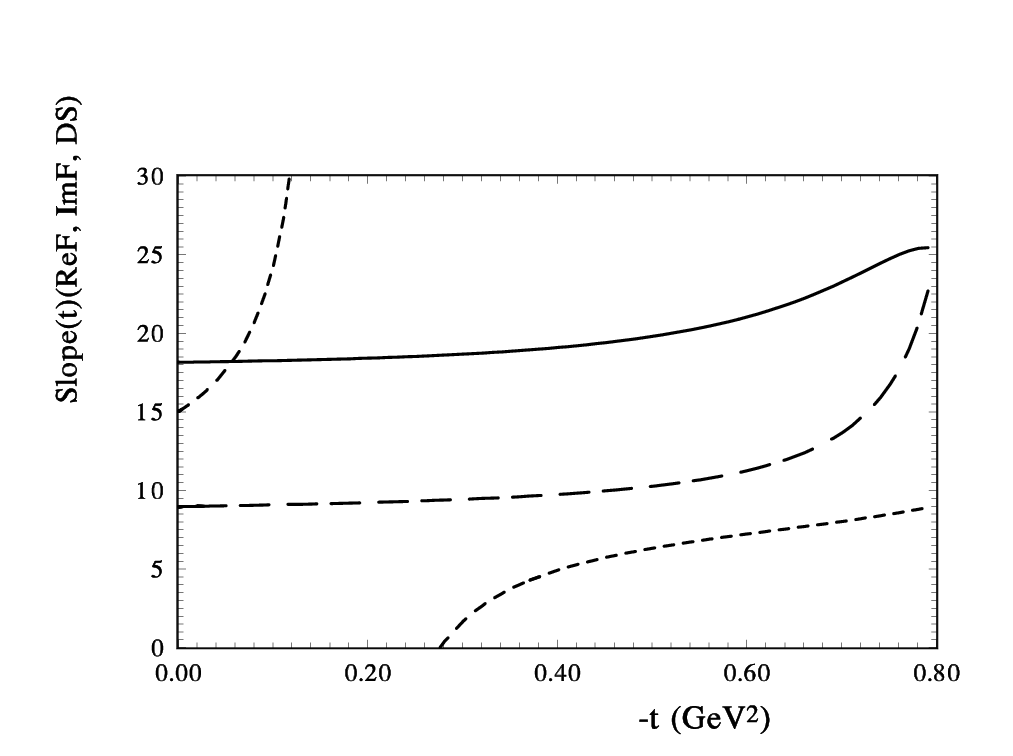}
\caption{  The slopes of the Born term of the scattering amplitude [top] and the uniterized
 scattering amplitude [bottom] ( long dashed line -the slope of the imaginary part, the short-dashed line  -
 the slope of the real part and the solid line - the slope of the differential cross sections.
 }
\end{figure}

\subsection{The real part of the elastic scattering amplitude}

  It is interesting that the form of the real part of the hadron elastic scattering amplitude
    is similar to its imaginary part. Of course, they are not proportional to each other as their connection has to satisfy the dispersion relations \cite{CS-PRL} which require, for example, the changing size of the real part at sufficiently small momentum transfer.
     In fact, in Fig. 14, we see that the real part changes its sign   essentially earlier than the position of the diffraction minimum.
   At LHC energies this happens in the area of momentum transfer approximately equal to $0.2$ GeV$^2$. But before that, we can see the crossing point at $|t| \sim 0.06$ GeV$^2$.
   Likely, the behavior of the imaginary part the slope of the real part
   at very small momentum transfer is also practically proportional to the size of the total cross section at different energies and grows at larger momentum transfers. It can be see that the real part
    at LHC energies has the negative maximum at approximately $|t|=0.3$ GeV$^2$ situated near the diffraction minimum. Hence, it  essentially impacts  the form and size of the diffraction minimum in the differential cross sections.

  Especially the size and energy dependence of the real part of the hadronic amplitude impact  the differential cross sections
 in the CNI region. The model descriptions of the differential cross section at a small momentum transfer and  at low energies
 are presented in Figures 5, 15-17. 
  These figures show the experimental data with high precision and  only statistical errors, which were taken into account
  in out fitting procedure.
  Obviously,  the model reproduces the experimental data very well in a wide energy region.


    One of the origins of the nonlinear behavior of the differential cross sections
    may be
    the different  $t$ dependencies of the imaginary and real parts of the scattering amplitude.
    In most part, in  different approaches it is supposed that this $t$ dependence is the same
    for both parts of the scattering amplitude.
    It should be noted the importance of  determining the size of the real part of the scattering
    amplitude  was emphasized in many work of Andre Martin. 
       If at the LHC the value of $\rho(s,t)$ is measured at high precision,
       it  gives the possibility to check up
    the validity of the dispersion relations \cite{Rev-LHC}.

     In the analysis of the experimental data \cite{T8a}
  two cases  were considered.  One is the so-called "central" case, in which the ratio of the real to imaginary
    parts of the scattering amplitude is independent of momentum transfer or slightly decreases.
    The other, the so-called "peripheral" case, takes into account the assumption that $\rho(s,t)$ grows with
    momentum transfer. Really, the last case contradicts the dispersion relations; hence, it has non-physical motivation.

      Really, as the scattering amplitude has to be an analytic function of its kinematic
        variable, let us take the energy dependence of the scattering amplitude through the complex
      $\hat{s} = s e^{-i\pi/2}$, and it must satisfy the dispersion relations.


     For simplicity, very often they use the so-called  local or derivative dispersion
     relations (see, for example, \cite{Bronzan}) 
     to determine the real part of the scattering
 amplitude.
     For example, the COMPETE Collaboration used
  \begin{eqnarray}
 &&Re F_{+}(E,0) =   \\ \nonumber
  &&(\frac{E}{m_{p}})^{\alpha} tan[\frac{\pi}{2} (\alpha-1+E\frac{d}{dE}]
  Im F_{+}(E,0)/(\frac{E}{m_{p}})^{\alpha}.
  \end{eqnarray}
   A different form of the derivative dispersion relation was taken as \cite{Roy-DR}
     \begin{eqnarray}
  &&Re F_{+}(E,0) =  \nonumber \\
  && (\frac{\pi}{ln(s/s_{0}}) \frac{d}{d\tau}[\tau Im F_{+}(s,t)/Im_{+}(s,t=0)] 
 ImF_{+}(s,t=0),
 \label{Roy-DR}
   \end{eqnarray}
   where $ \tau= t (ln(s/s_{0})^{2}$ and as $s \rightarrow \infty$.
     To satisfie these relations, the scattering amplitude has to be
      a unified analytic function of its kinematic variables connecting different reaction channels.

\begin{figure}[b]
\includegraphics[width=65mm]{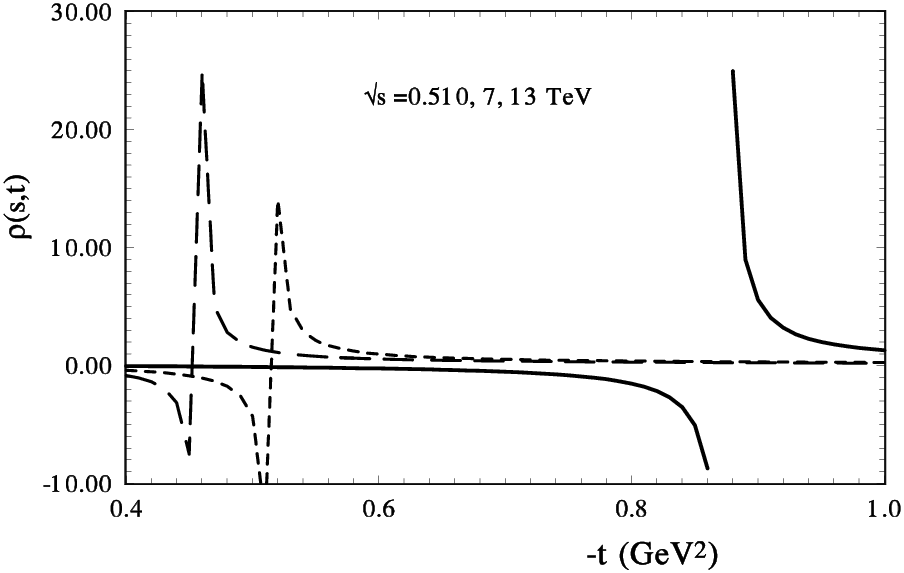}
\caption{ The size of the $\rho(s,t)$ - ratio of the real to imaginary part of the $pp$ scattering amplitude
is calculated in the model at the energy  $\sqrt{s} =  0.51, 7, 13$ TeV
 (solid, short dashed and dashed line respectively)   depending on $s$ and $t$ }.
\end{figure}

           In most cases, one assumes that the real part is taken  proportional to the imaginary part of the scattering amplitude.
            So the slopes
           of both parts are equal to each other.
           There is also some unusual assumption about the growth
           of the real part of the scattering amplitude at a small momentum transfer relative to its imaginary part
           (the so-called  the peripheral case \cite{T8a}).
           Obviously, both assumptions do not satisfy the dispersion relations, especially the last one.

 Let us examine the origins of the  
  complicated $t$ dependence of the real part.
  Take the scattering amplitude in the form
   \begin{eqnarray}
   F(s,t) \ = \ h s^{\Delta} e^{B t Ln(\hat{s})} .
   \label{cExp}
\end{eqnarray}
  where the complex $\hat{s}$ is used only in the exponential.
  In this case, the real part is negative (in Fig. 13, it is shown by the short dashed line)
  and essentially differs from the behavior of the imaginary part (long-dashed line).
   Now let us use  $\hat{s}$ in all parts of the scattering amplitude
   \begin{eqnarray}
   F(s,t) \ = \ h \hat{s}^{\Delta} e^{B t \ln(\hat{s})} .
   \label{sscExp}
\end{eqnarray}

  The dispersion relations lead to the fact that the slope of the real part of the scattering amplitude
           must be larger than the slope of the imaginary part.
           For example, if the imaginary part of the spin-nonflip hadron elastic scattering amplitude takes a simple exponential form $Im F_{+} \sim h e^{Bt}$, then from eq.(\ref{Roy-DR}) we have
           that the  real part of the $F_{+}$ is $ ReF_{+} = (1.+Bt)e^{Bt}$. Hence, it has zero in the
           region of momentum transfer around $-0.1 - -0.15$ GeV$^2$

   At last, but not least, it should  be noted that the unitarization procedure has  a strong  influence on the $t$ dependence of the real part of the scattering amplitude. For example, take the standard eikonal form of  unitarization. If the Born term is taken in the ordinary exponential form (eq.\ref{sscExp}), then the imaginary part has a constant slope (see Fig. 15 [top]) and the real part has the first zero at a small momentum transfer. After eikonalization both slopes of the imaginary  and real parts of the scattering amplitude  have a strong $t$ dependence (Fig. 15) [bottom]).

\section{ The value of $\rho(s,t)$  and Odderon contributions  }

    In Fig. 16, the ratio $\rho(s,t)$ of the real to imaginary part  of
    the hadron elastic scattering amplitude is presented for different energies.
    Such a complicated structure of   $\rho(s,t)$ is determined by the changes of the sign of the real and
    imaginary parts of the scattering amplitude. At a small momentum transfer, the size of $\rho(s,t)$ is small
    as the real part changes its sign. Contrary, when the imaginary part changes its sign, the size of $\rho(s,t)$ grows very faster. The energy dependence of $\rho(s,t)$ is due to the movement of the position of the diffraction minimum,
     and hence to the energy dependence of the imaginary part of the scattering amplitude.

          At high energies and in the region of small momentum transfer the difference
          between the $pp$ and $p\bar{p}$ differential cross sections
          comes in most part from the CNI term, as  the real part of the Coulomb amplitude has
         a different sign in these reactions.
       In the standard fitting procedure, one neglects the $\alpha_{em}^2$ term  
       and then the equation takes the form:
\begin{eqnarray}
d\sigma/dt =&& \pi [ (F_{C} (t))^2\!+\! (\rho(s,t)^2 + 1) (Im F_{N}(s,t))^{2})  \nonumber \\
+ && 2 (\rho(s,t)+ \alpha \varphi(t)) F_{C}(t) Im F_{N}(s,t)],
\label{ds2b}
\end{eqnarray}
where $F_{C}(t) = \mp\ 2 \alpha G^{2}(t)/|t|$ is the Coulomb amplitude (the upper sign is for $pp$, the lower sign is for $p\bar p$) and $G^2(t)$ is  the  proton electromagnetic form factor squared;
$ReF_N(s,t)$ and $ImF_N(s,t)$ are the real and imaginary parts of the hadron amplitude;
$\rho(s,t) = Re F_N(s,t) / Im F_N(s,t)$.
Formula (\ref{ds2b}) is often used for the fit of experimental  data in getting hadron amplitudes and the Coulomb-hadron phase
in order to obtain the value of $\rho(s,t)$.
 One can clearly see   the impact of the value of $\rho(s,t)$
on the form and energy dependence of the differential cross sections.


\begin{figure}
%
\includegraphics[width=0.4\textwidth]{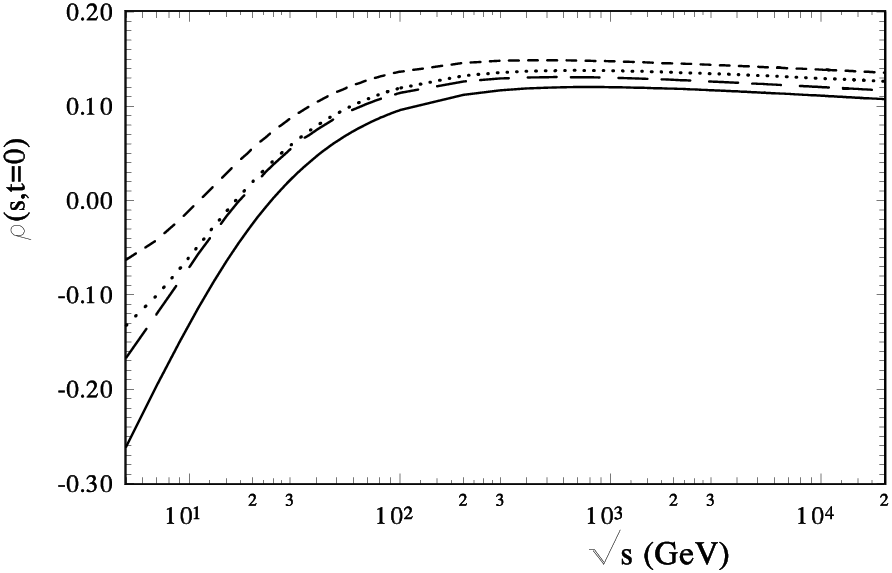}
\caption{
      $\rho(s,t=0)$: dashed line - $p\bar{p}$ scattering, solid line -$pp$ scattering;
       the same but without the second Rejions contributions (tiny dashed - $p\bar{p}$ and long dashed $pp$ scattering).
  }
\label{Fig_16}
\end{figure}


\begin{figure}
\includegraphics[width=0.4\textwidth] {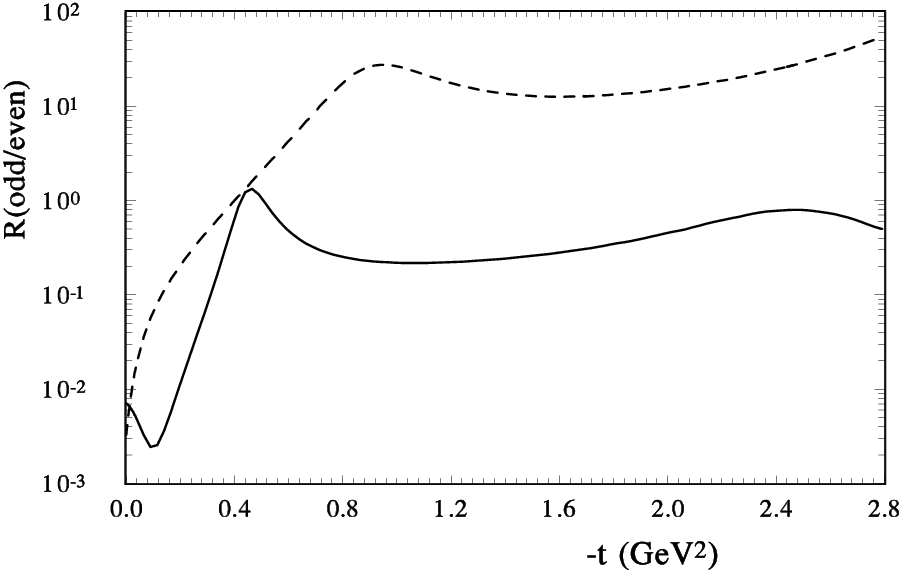}
\vspace{0.5cm}
\caption{  The ratio $R(s,t)$ of the cross-odd part of the scattering amplitude of elastic proton-proton scattering to its cross even part  
at 13 TeV (dashed line) and at 9.26 GeV (solid line).
 }
\end{figure}


\begin{figure}
\includegraphics[width=0.4\textwidth] {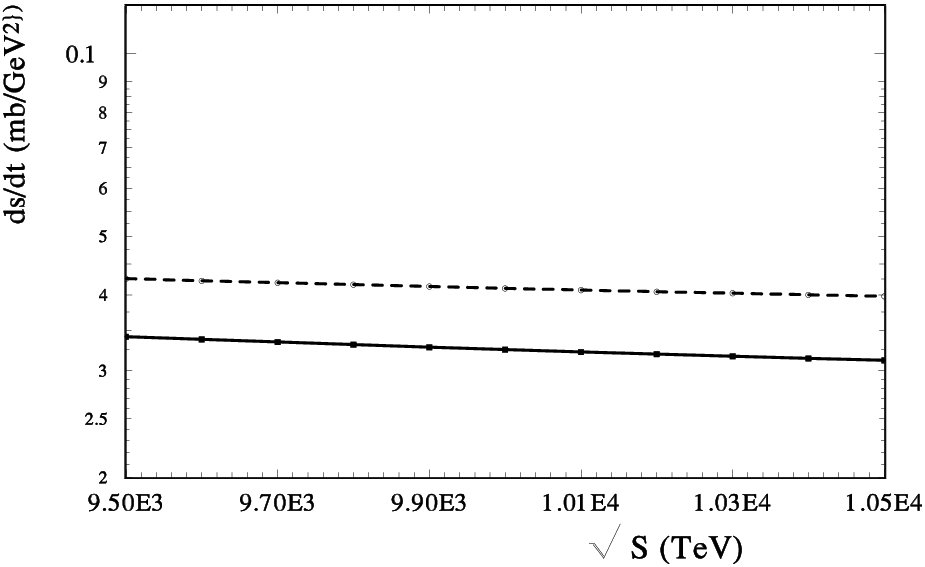}
\includegraphics[width=0.4\textwidth] {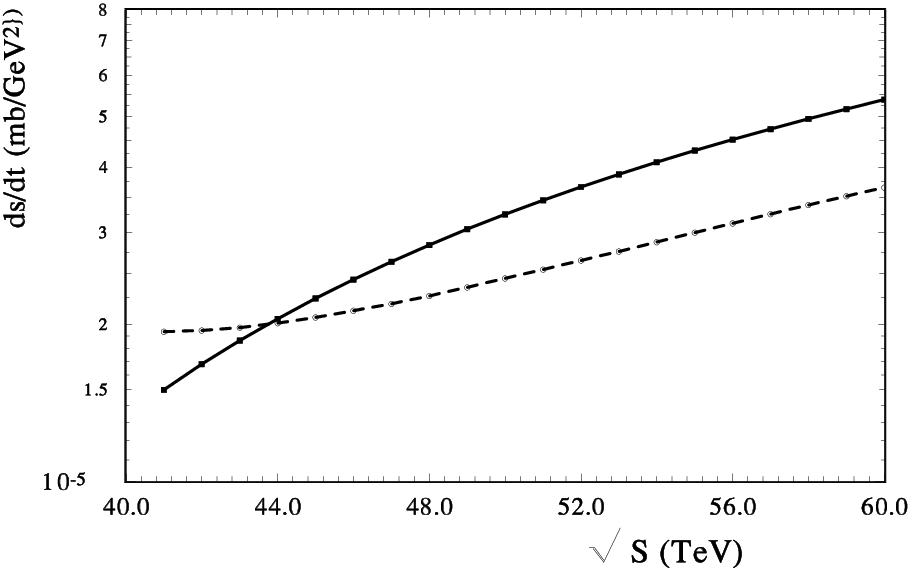}
\caption{  The contributions of the Odderon part to the differential cross sections
in the region of the dip a)[top] $-t=0.45$ GeV$^2$ ; b) [down]  $-t=1.45$ GeV$^2$.
 }
\end{figure}

  The Odderon amplitude  has a large real part compared to it imaginary part.
   Hence, the odderon contribution changes the size and $t$ dependence of the real part of the full amplitude.
  In a recent paper \cite{Khoze-Odd24}  different cases (with and without Odderon contributions) were analysed. 
  It was  noted that the effect of incorporating the Odderon
  becomes notably significant when analysing specific subsets of data.
  It is  remarkable that the authors note "we will get too large $\rho^{p\bar{p}}$ at
   $\sqrt{s} \sim 541$ \ GeV in disagreement with the data UA4/2. 
   As a result, they restored an old problem of the value of $\rho(s,t)$ at $\sqrt{s} \sim 540$ GeV.
   However, as was noted in the introduction,  this problem strongly depends on the
   form of the non-linear slope and can be solved  in \cite{SelYF92,Sel-UA42}.
   Now, in the present model, the Odderon amplitude essentially
    decreases as $t \rightarrow 0$. The value of $\rho(\sqrt{s} = 541 \rm{GeV},t=0) =0.122$,
    which coincides with the result of the UA4/2 Collaboration.


\begin{figure}[b]
\includegraphics[width=0.4\textwidth] {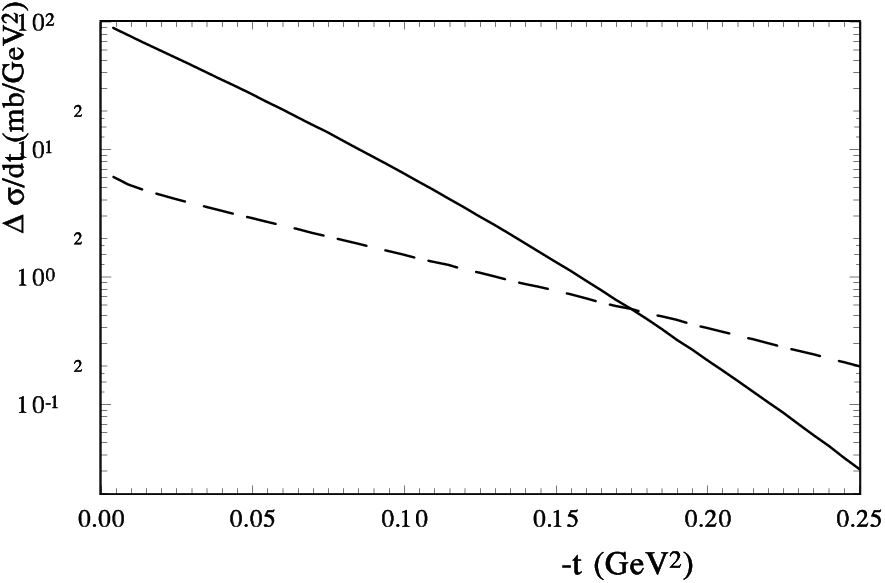}
\caption{The contributions  $\Delta(d\sigma/dt)$ of the anomalous term with a large slope to the differential cross sections
 of $pp$ scattering at $\sqrt{s}=19$ GeV (solid line) and at $\sqrt{s}=13$ TeV.
 }
\end{figure}

     In Fig. 18, the ratio of the cross-odd part of the scattering amplitude in the HEGS model of elastic proton-proton scattering
       to its cross even part at 13 TeV  and at 9.26 GeV is presented.
        It is seen that this ratio is small at 13 TeV, except for the position of the diffraction minimum.
        At low energy, this ratio is small at   small t  but increases essentially  in the region of a large momentum transfer.

    However, the Odderon contribution is very important at the diffraction minimum.
    The real part of the scattering amplitude, which is positive at $t=0$ and $\sqrt{s}> 30$ GeV,
     changes its sign   at  larger $t$, thus corresponding to the dispersion relations
     but, in  any case, it fulfills the diffraction dip in the differential cross sections.
     Figure 17  shows the sizes of the differential cross sections in the region of
     the diffraction minimum with and without the Odderon contributions at
     $\sqrt{s}$ around $53$ GeV and at $10$ TeV.

\section{New effects in diffraction elastic scattering
                        at small angles}

   In the fitting procedure of the experimental data, only statistical errors were taken into account.
As the systematic errors are mostly determined by indefiniteness of luminosity,
they are taken into account as an additional normalization coefficient.
 This method essentially decreases the space of a possible form of the scattering amplitude.
 This allowed us to find
the manifestation of  small effects at 13 TeV experimental data for the first time  [19-21].
   Our further researches with taking into account a wider range of experimental data confirm the existence of such new effects.
 We determined the new anomalous term with a large slope as
\begin{eqnarray}
 f_{an}(t)= && i h_{0an} \ln{(\hat{s}/s_{0})}/k (1+ h_{1an} \ln{(\hat{s}/s_{0})}/k)  \nonumber  \\
  && \exp[-\alpha_{an} (|t|+(2t)^2/t_{n})\ln{(\hat{s}/s_{0})}] \   G^2_{em}(t); 
  \label{fan}
\end{eqnarray}
where $h_{an}$ is the constant determining the size of the anomalous term with a large slope - $ \alpha_{an}$;
  $G^2_{em}(t)$ is the electromagnetic form factor,
  which was determined 
   from the 
  GPDs \cite{GPD-PRD14}, and
  $ \ k=\ln(13000^2  \ {\rm GeV}^{2}/s_{0}) \ $  is  introduced for normalization of $h_{an}$ at $13$ TeV,
 and  $t_{n}=1$ GeV$^2$ is the normalization factor. 
    This form adds only two additional fitting parameters,
 and    this term is supposed to grow with energy of order $ \ln{(\hat{s}/s_{0})}$.
  The  analysis of the data in a wide energy region gives
   a complicated energy dependence which is reflected in eq.(\ref{fan})
      The term has a large imaginary part
and a small real part determined by the complex $\hat{s}$.
  It is remarkable that the anomalous slope is determined with high accuracy  $ \alpha_{an}=0.513\pm0.004$ GeV$^{-2}$
  from the analysis of the whole sets of experimental data.


\begin{figure}
\begin{center}
\includegraphics[width=0.4\textwidth] {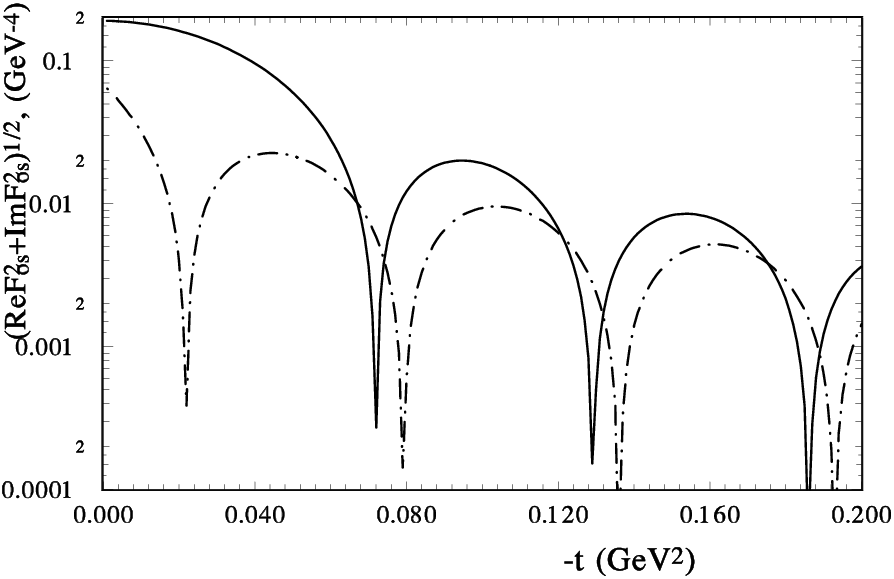}
   \includegraphics[width=0.4\textwidth] {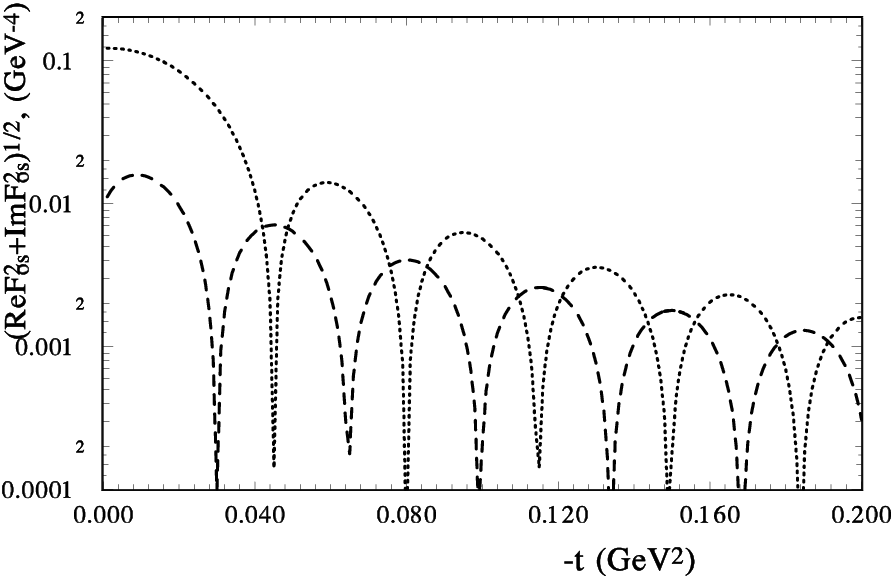}
\end{center}
\vspace{1.cm}
\caption{    The absolute value of the oscillation term $(Re^{2}F_{os} + Im^{2}F_{os})^{1/2}$ at $\sqrt{s}=13$ TeV
 (solid line $pp$ scattering; dashed points line -  $p\bar{p}$ scattering ).
  and at    $\sqrt{s}=540$ GeV (dotted line $pp$ scattering; dashed  line -  $p\bar{p}$ scattering)
%
  }
\label{fig:20}       
\end{figure}

\begin{figure*}
\begin{center}
\includegraphics[width=0.4\textwidth] {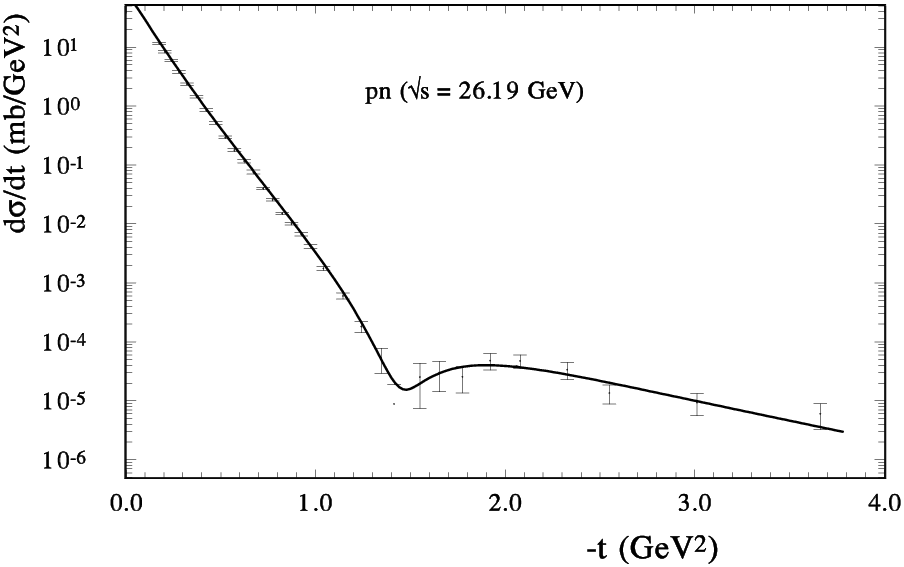}
\includegraphics[width=0.4\textwidth] {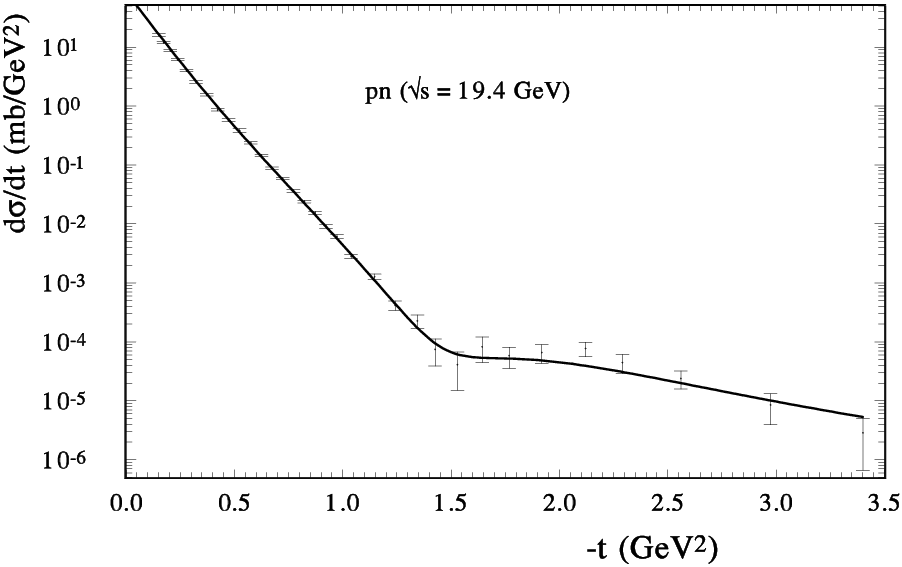}
\end{center}
\vspace{0.5cm}
\caption{The HEGS model calculation of $d\sigma/dt$ of  $pn $ scattering \\
[left] at $\sqrt{s} =26$ GeV; [right] $\sqrt{s} =19.4$ GeV.
   }
\end{figure*}

  It is proportional to electromagnetic form factors, and the analysis of the experimental data above
 $6$ GeV gives the size of the constant $h_{an}=1.13 \pm 0.04$. 
  It is a little less than obtained in the fit of only high energy data but now it has a more complicated energy dependence and a very small error.
   It impacts  the differential cross sections at small $t$ and  the size of  $\sigma_{tot}(s)$.
   However, it practically does not influence the value of $\rho(s,t)$ as the imaginary and real parts of the scattering amplitude
   grow proportionally.
   The contributions   of the anomalous term with  a large slope to the differential cross sections
 of $pp$ scattering -
   $$\Delta(d\sigma/dt= d\sigma/dt|_{with} -  d\sigma/dt|_{without})$$
   are shown in Fig. 20 at $\sqrt{s}=19$ GeV (solid line) and at $\sqrt{s}=13$ TeV (dashed line).
  It can be seen that the contribution of the anomalous term disappears after
   $-t>0.2$ GeV$^{2}$. At low energies its relative contribution
   is less but remains visible and decreases with $t$ slowly.

Our method helps us to find a small oscillation effect in the differential cross section at small momentum transfer \cite{Osc13}.   Such oscillation can be determined by an additional oscillation term in the scattering amplitude \cite{Selppap}. 
 The theoretical predictions of some oscillations in the differential cross section
 has a long story, starting from  one of the first paper \cite{Anselm} taking into account a multi pomeron exchange and
 the papers basing on the violation of the Pomeranchuk theorem (for example, \cite{AKM})   up to a recent paper \cite{Ryskin}.
 In  our fitting procedure
  the oscillatory function is
%
\begin{eqnarray}
 f_{osc}(t)= &&  i  h_{0osc} (1 \pm i  h_{1osc})  \\ \nonumber
 && (\ln{(\hat{s}/s_{0})}/k +h_{s}/\hat{s})  \   J_{1}(\tau)/\tau \ A^{2}(t), \\
  && \tau = \pi \ (\phi_{0}-t)/t^{\prime}_{0}; \nonumber
  \label{osc}
\end{eqnarray}
here $J_{1}(\tau)$ is the Bessel function of the first order;
     $t_{0}=1/[a_{p}/(1+2 \ln^{2}{(\hat{s}/s_{0})}/k)]$, where $a_{p}=17.15$ GeV$^{-2}$ is the fitting parameter
     that leads to the AKM scaling on  $\ln{(\hat{s}/s_0)} $;
     $A(t)$ is the gravitomagnetic form factor, which was determined from the
  GPDs \cite{GPD-PRD14}, and
   $h_{osc}$ is the constant that determines the amplitude of the oscillatory term with
   the period determined by $\tau$.
 This form has only a few additional fitting parameters and allows one to represent
 a wide range of  possible oscillation functions.
  The phase $\phi_{0}$ is obtained near zero for $pp$ scattering
  and  it is slightly  small and negative for  $p\bar{p}$ scattering,
  so it has a different sign for $pp$ and $p\bar{p}$ scattering.
 The inclusion  of the  $p\bar{p}$ elastic scattering  data  in the fitting procedure shows
 that  part of the oscillation function changes  its sign for the crossing reactions.
 As a result, the plus sign is related to  $pp$ and minus with  $p\bar{p}$ elastic scattering.
   Hence, this part is  the crossing-odd  amplitude, which has the same simple form for $pp$ and
  $p\bar{p}$ scattering  only with a different sign.

\begin{figure*}
\begin{center}
\includegraphics[width=0.4\textwidth] {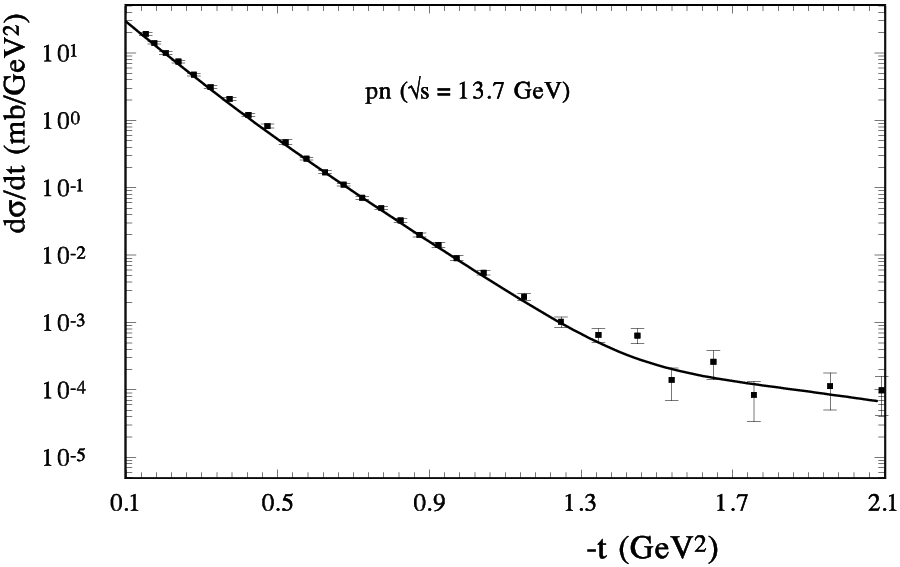}
\includegraphics[width=0.4\textwidth] {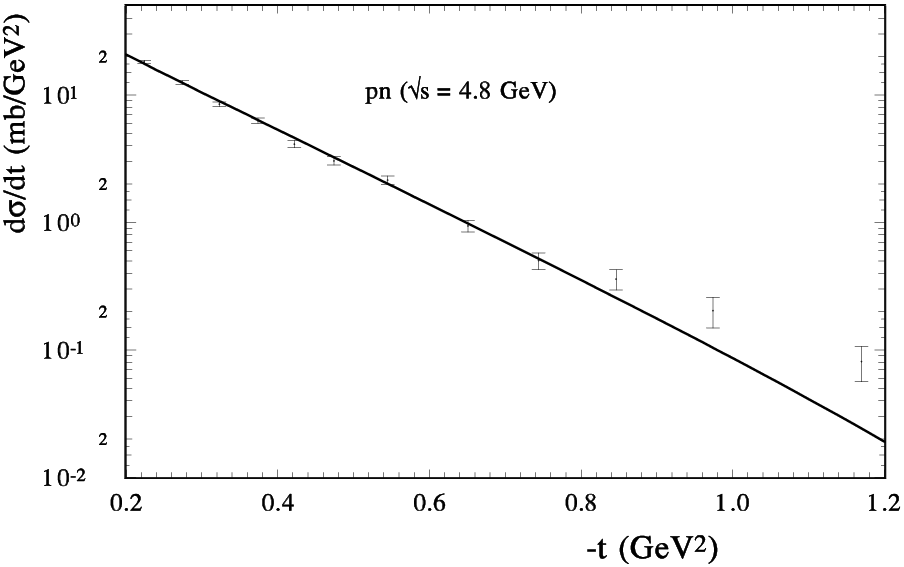}
\end{center}
\caption{The HEGS model calculation of $d\sigma/dt$ of  $pn $ scattering \\
[left] at $\sqrt{s} =13.7$ GeV; [right] $\sqrt{s} =4.8$ GeV.
   }
%
\end{figure*}

   The wider energy region used in this analysis allows one to reveal the logarithmic energy dependence of the
   oscillation term.
    The constant (size) of the oscillation function was determined from experimental data
    at high energies (all data above $500$ GeV)
    as $h_{osc}^{c}=0.270\pm0.007 \ {\rm GeV}^{-2}$,  whereas in our analysis of the whole sets of the data $\sqrt{s}> 3.6$ GeV.
    the $h_{osc}^{c}=0.232\pm0.009 \ {\rm GeV}^{-2}$
 The size of $  h_{osc}$ is obtained as  is smaller than obtained in the only high energy data; however, the error is decreased essentially.
   Perhaps,  this reflects a more complicated form of the  energy dependence obtained now.
    Note, despite  the logarithmic growth of the oscillation term, its relative contribution  decreases
    as the main scattering amplitude grows as $\ln^{2}(s)$.

    The oscillation term is represented in Fig. 21 at different  energies   $\sqrt{s}=13$ TeV and $\sqrt{s}=540$ TeV
    for the proton-proton and proton-antiproton scattering.
    It can be seen that the amplitude of the oscillation term increases at high energies, but the period of oscillations
     decreases with growing energy. Especially note that the period of oscillation is essentially less for $p\bar{p}$ than for $pp$ scattering. This can explain the difference  which was noted in \cite{Grafstr}, in the behaviour of oscillations of the differential cross sections at $\sqrt{s}=540$ GeV,
    which were found in \cite{Gauron-97} and at  $\sqrt{s}=13$ TeV \cite{Osc13}.
    Of course, the amplitude of oscillations is weak, so it is very difficult to identify it by experiment.
    However, our analysis of a wide energy region of experimental data gives a small error in the definition
    of the parameter determining the size of oscillations.  
    This shows that further meticulous work is required, both from the theoretical and experimental points of view.

\section{Proton-neutron elastic scattering}

   We take in our analysis $24$ sets of the proton-neutron experimental data
   from $\sqrt{s} > 4.5$ GeV up to maximum energy $\sqrt{s} > 27.5$ GeV  where they
  correspond to experimental data with a total number of experimental points $N=526$.
    As in the case of $pp$ scattering, we include only
    statistical errors in our fitting procedure
      taking into account the systematic errors as additional normalization of a separate set. We take the amplitudes of the model obtained
    in the case of $pp$ and $p\bar{p}$ scattering and fix the parameters of
   the main terms and two anomalous terms, except for the terms of the second Reggions. The electromagnetic amplitudes
    are removed in this case.
      As a result, in the $pn$ case, only the parameters of the second effective  Reggion are obtained from the fitting procedure.
      Only in the Pomeron-neutron vertex $ F_{Pom2}^{Born} n$ is taken in the standard dypole form with free $\Lambda_{pom-n}$.
    Its  value is obtained  to be slightly above the standard electromagnetic $\Lambda_{pom-n}^2 =0.82$ GeV$^{2}$.
     The obtained $\sum_{i,j} \chi^2_{i,j}=567$.
    The corresponding obtained description of the differential cross section is represent
    in Figs. 4, 21, 22.  It is clear that the HEGS model  works very well in the case
    of proton-neutron elastic scattering.
      In Fig. 22, the differential cross section of $pn$ elastic scattering is represented for a large momentum transfer
      at energies $\sqrt{s}=26.19$  GeV and  $\sqrt{s}=19.4$ GeV.
      At  $\sqrt{s}=26.19$  GeV  (Fig. 22[left]) the diffractive dip-bump structure is clearly  represented
      whereas at $\sqrt{s}=19.4$ GeV (Fig. 22[right]) this structure is being leveled out
      and practically disappears at more lower energies
      (see Fig. 23).
     As in the case of $pn$ scattering, there are
    many experimental data at small momentum transfer, where the contributions
     of our two anomalous terms are important; the successful description of the differential cross section
     confirms the existence of these terms.

\begin{figure*}
\begin{center}
\includegraphics[width=0.4\textwidth] {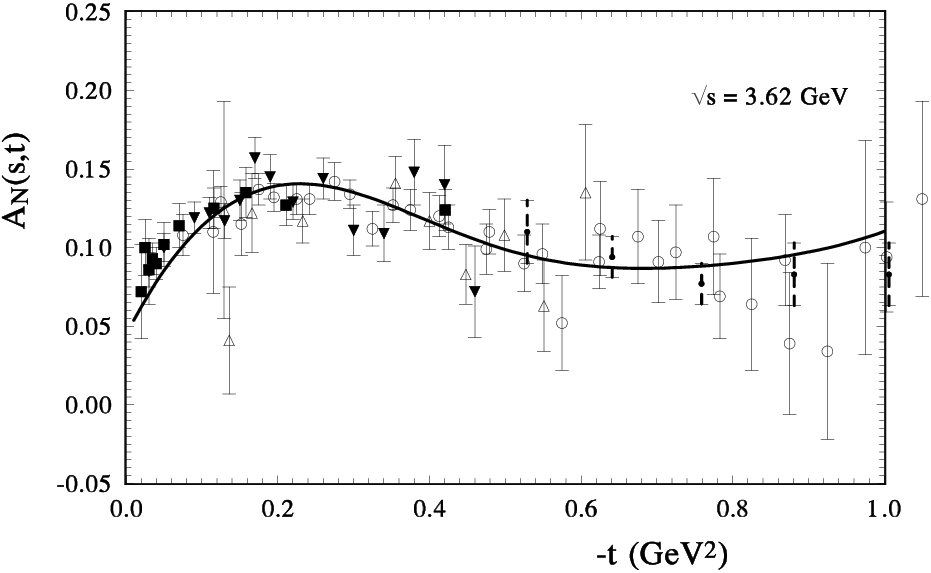}
\includegraphics[width=0.4\textwidth] {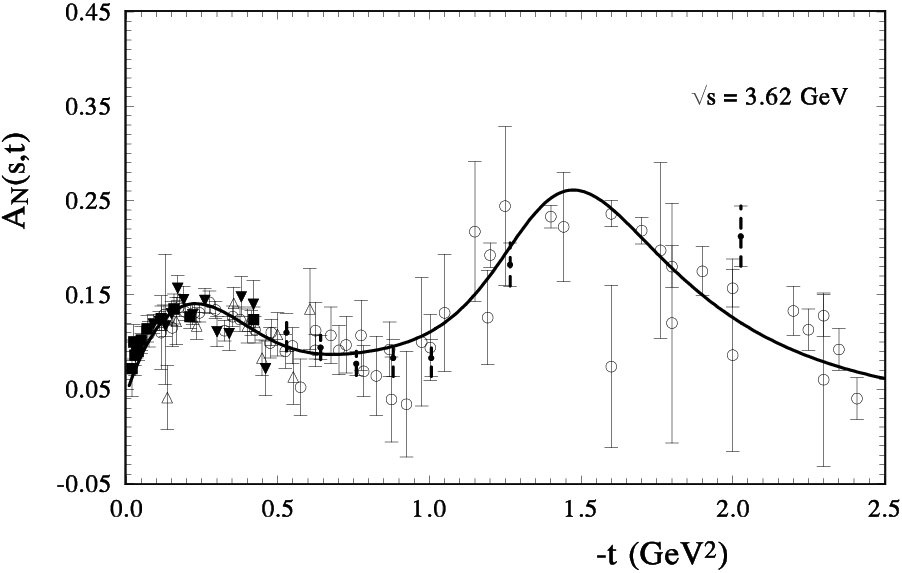}
\end{center}
\caption{The analyzing power $A_N$ of pp - scattering is calculated:
[left] at $\sqrt{s} =3.62$  GeV; (small $t$), 
[right] (larger $t$). 
 }
%
\label{fig:22}       
\end{figure*}

\begin{figure*}
	\begin{center}
		\includegraphics[width=0.4\textwidth] {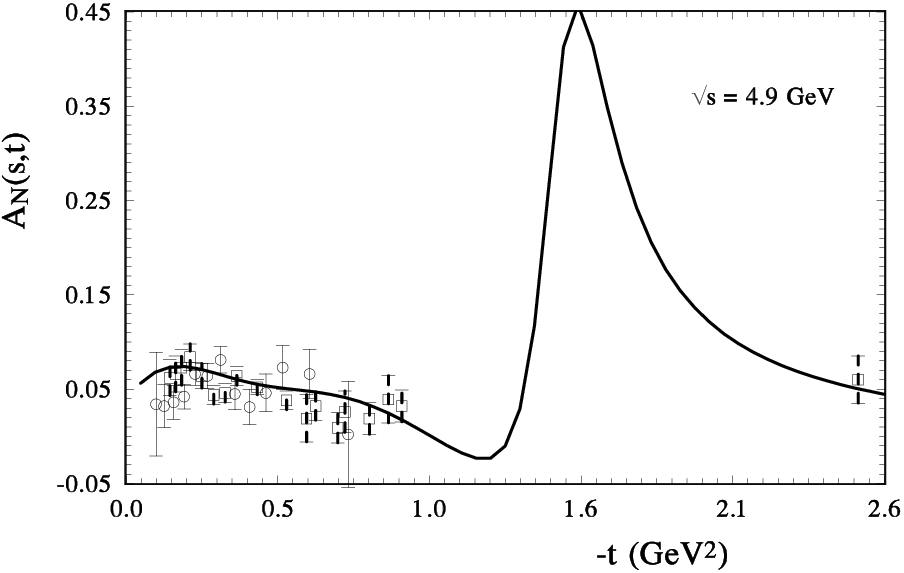}
		\includegraphics[width=0.4\textwidth] {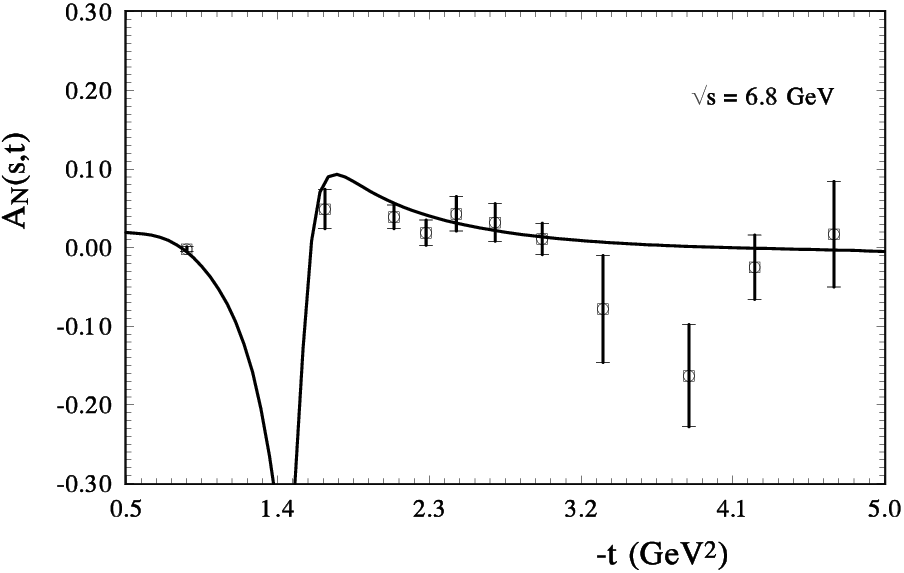}
	\end{center}
	\caption{The analyzing power $A_{N}$ of pp - scattering	calculated:
    		[left]  at $\sqrt{s} = 4.9 \ $GeV,  (the experimental data \cite{HEP-data}), and
		[right]     at $\sqrt{s}= 6.8 \ $GeV (points - the experimental data \cite{HEP-data}).
	}
	\label{fig:23}       
\end{figure*}

\begin{figure*}
	\begin{center}
		\includegraphics[width=0.4\textwidth] {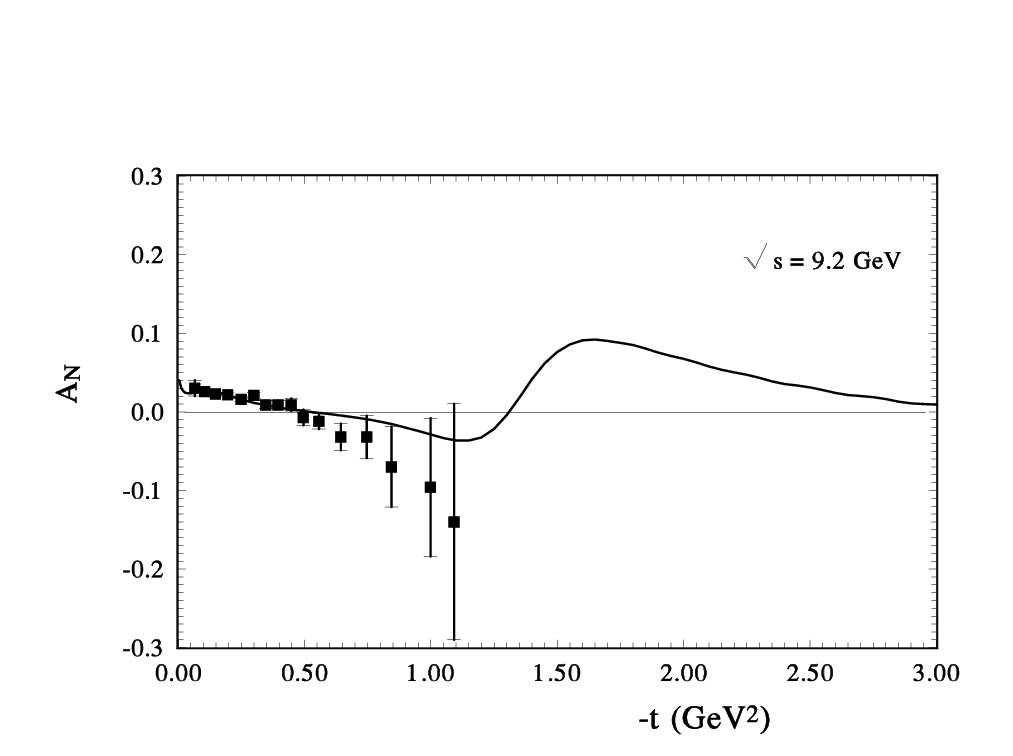}
		\includegraphics[width=0.4\textwidth] {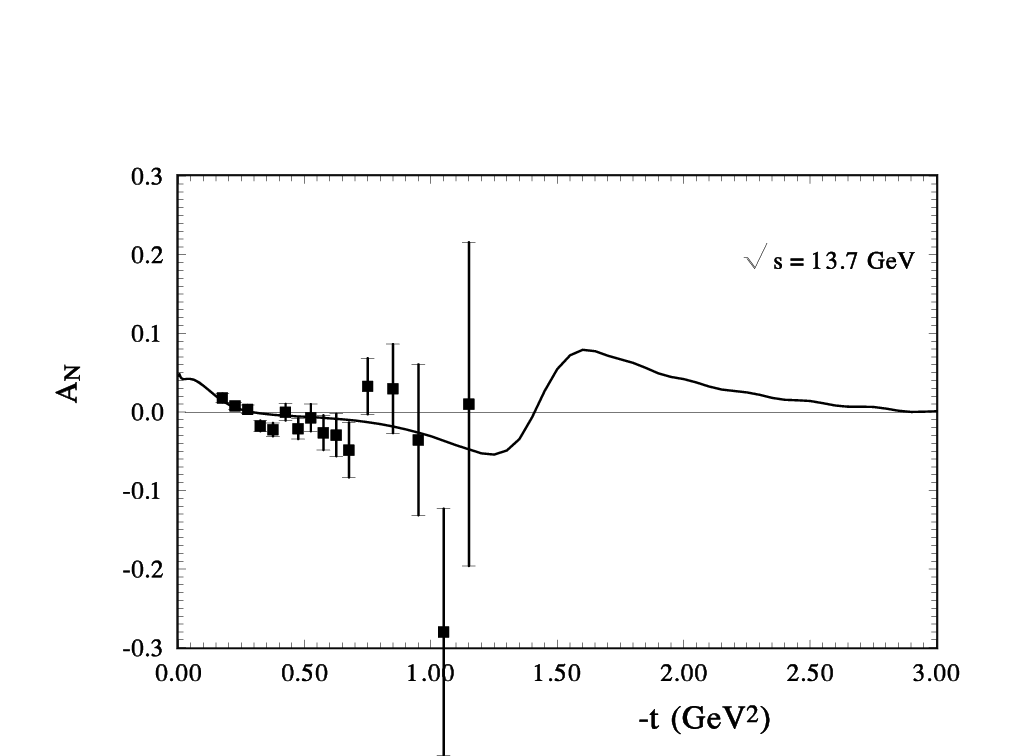}
	\end{center}
	\caption{The analyzing power $A_N$ of pp - scattering
		calculated:
		[left]  at $\sqrt{s} = 9.2 \ $GeV  (the experimental data \cite{HEP-data}) and
		[right]     at $\sqrt{s}= 13.7 \ $GeV (points -
		the experimental data \cite{HEP-data}).
	}
	\label{fig:24}       
\end{figure*}

\begin{figure*}
\begin{center}
\includegraphics[width=0.4\textwidth] {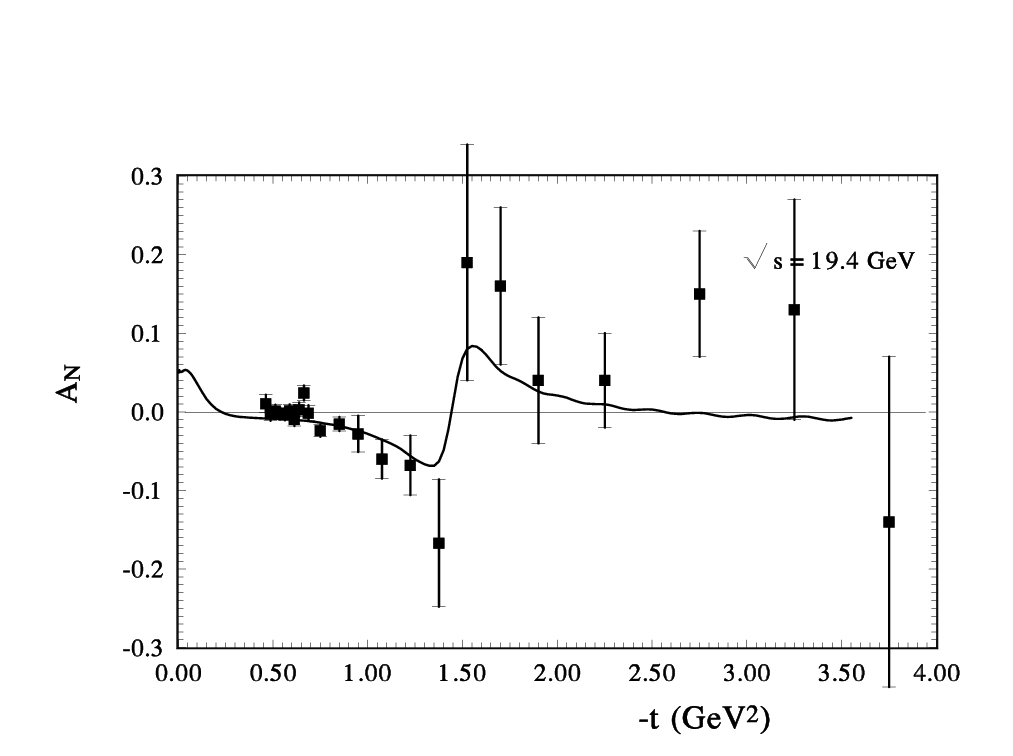}
\includegraphics[width=0.4\textwidth] {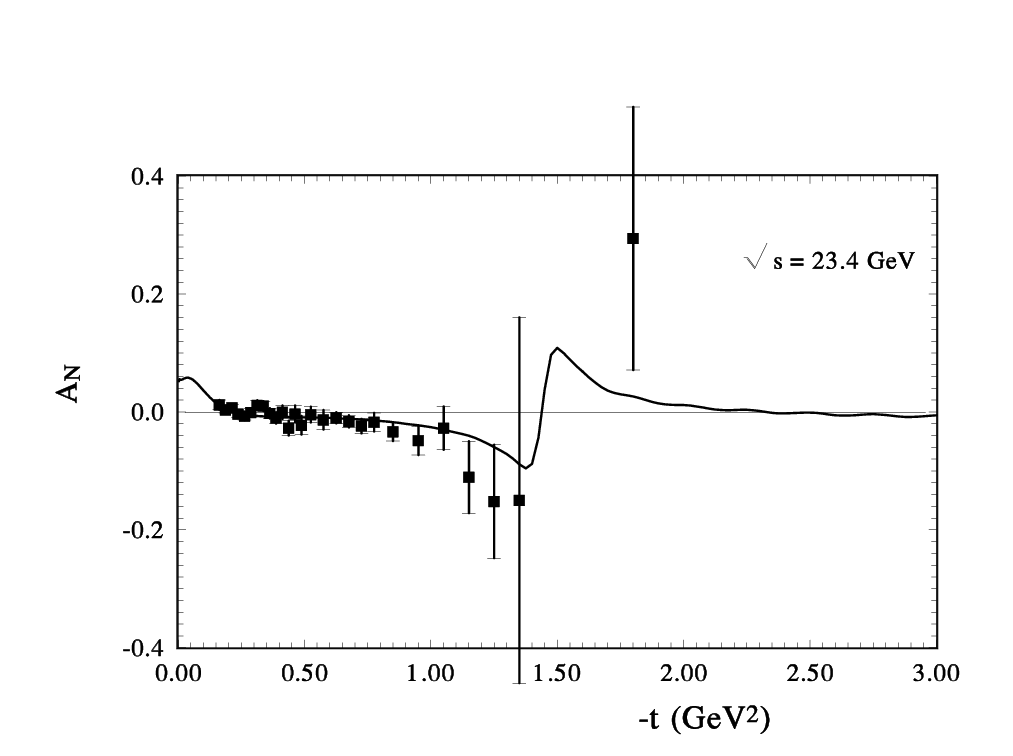}
\end{center}
\caption{The analyzing power $A_N$ of pp - scattering
      calculated:
   [left]  at $\sqrt{s} = 19.4 \ $GeV  (the experimental data \cite{HEP-data}), \\
   and
   [right]     at $\sqrt{s}= 23.4 \ $GeV
       (points - the existing experimental data \cite{HEP-data})
          }
\label{fig:25}       
\end{figure*}

\section{Polarization effects in proton-proton elastic scattering}
      In the  Regge limit $t_{fix.}$ and $s \rightarrow \infty$  one can write the
       Regge-pole contributions to  the helicity amplitudes in the $s$-channel as
\begin{eqnarray}
    \Phi^{B}_{\lambda_1, \lambda_2,\lambda_3, \lambda_{4}} (s,t) \sim&& \\ \nonumber
   && \sum_{i} g^{i}_{\lambda_1, \lambda_2}(t)  g^{i}_{\lambda_3, \lambda_4}(t) 
   [\sqrt{|t|}]^{|\lambda_1- \lambda_2|+|\lambda_3- \lambda_4|} \\ \nonumber
   && (\frac{s}{s_0})^{\alpha_{i}} (1 \pm e^{-i \pi \alpha_{i} }).
\end{eqnarray}

The  corresponding  spin-correlation values are presented in eq.(\ref{AN}).

 Neglecting  the $ \Phi_{2}(s,t)- \Phi_{4}(s,t)$ contribution, the spin correlation parameter $A_{N}(s,t)$
   can be written taking into account the phases of  separate spin non-flip and spin-flip amplitudes
   as  $\varphi_{nf}(s,t), \varphi_{sf}(s,t)$   the analyzing power is
  \begin{eqnarray}
  A_{N}(s,t)  =  && -\frac{4 \pi}{s^2}
   [ (F_{nf}(s,t)| \ |F_{sf}(s,t)|     \\ \nonumber
   && sin(\varphi_{nf}(s,t)-\varphi_{sf}(s,t))/ \frac{d \sigma}{dt}.
   \label{ANs}
\end{eqnarray}
  It is clearly seen that despite the large spin-flip amplitude, the analyzing power can be near zero
  if the difference of the phases is zero in some region of momentum transfer.
  The experimental data at some point of the momentum transfer show the energy independence of
  the size of the spin correlation parameter $A_{N}(s,t)$. 
  Hence, the small value of the $A_{N}(s,t)$ at some $t$ (for example, very small $t$)
  does not serve as a proof that it will be small in other regions of momentum transfer.

 It is usually assumed  that the imaginary and real parts of the spin-non-flip
  amplitude have the exponential behavior with the same slope, and the
  imaginary and real parts of the spin-flip amplitudes, without the
  kinematic factor $\sqrt{|t|}$    \cite{sum-L},
   are proportional to the corresponding parts of the non-flip amplitude.
  That is not so as regards the $t$ dependence
  shown in Ref. \cite{soff}, where
  $F^{fl}_{h}$ is multiplied 
  by a special function
   dependent on $t$.
  Moreover, one mostly  takes  the energy independence of
  the ratio of the spin-flip parts to the spin-non-flip parts of the
  scattering amplitude.                         
All this is our theoretical uncertainty \cite{Cudpred-EPJA,M-Pred}.

  In \cite{G-Kuraev1,G-Kuraev2}, on the basis of  generalization of the constituent-counting rules of the perturbative QCD,  the proton current matrix elements $J^{\pm\delta\delta}_{p}$
   for a full set of  spin combinations  corresponding to the number
   of the spin-flipped quarks were calculated. This  leads to part of the spin-flip amplitude
 \begin{eqnarray}
              F_{h}^{sl}  \sim \sqrt{-t}/(\frac{4}{9} m_{p}^{2}) \  \sqrt{-t}/(\frac{4}{9} m_{p}^{2}) \
              \sqrt{-t}/(\frac{4}{9} m_{p}^{2}).
       \label{GKur-sflip}
\end{eqnarray}
  Hence, such an amplitude  gives  large contributions at large momentum transfer.

   Of course, at  lower energies we need to take into account
  the energy-dependent parts of the spin-flip amplitudes.
 So the form of the spin-flip amplitude is determined as
 \begin{eqnarray}
 F_{sf1}(s,t)=i h_{sf1} q^3 (1+q^3/\sqrt{\hat{s}}) 
  G_{em}^2 e^{2 t ln{\hat{s}}}
\end{eqnarray}


  We take the second part of the spin-flip amplitude in the form
  \begin{eqnarray}
  F_{sf2}(s,t)= i \sqrt{|t|} G_{em}^{2} (h_{5} +h_{6} 
  (1 + i h_{4})/ssc^2) e^{2 t ln{\hat{s}}}
\end{eqnarray}
  This  works in most part at low energies.

 The minimal energy at which the analyzing power $A_{N}(s,t)$ of pp - scattering
  examined in  the HEGS model is   $\sqrt{s}=3.62 $~GeV. 
  For our high energy model it is a very small energy.
 These results are shown in Fig. 24 at small $t$  [left] and at larger $t$ (right).
  These  model calculations are compared with the results of  four experiments.
  Obviously, we obtain a very good description of the experimental data.
  At these energies, the diffraction
  minimum is practically  overfull by the real part of the spin-non-flip amplitude and
  the contribution of the spin-flip-amplitude; however, the $t$-dependence of
the  analysing power is very well reproduced in this region of the momentum transfer.


  Our calculation for $A_{N} (t)$ is shown in Fig. 25 (left, right) at $\sqrt{s}=4.9 $~GeV and $\sqrt{s}=6.8 $~GeV.
  The description of the existing data is sufficiently good.
 Note that the magnitude and the energy dependence of this parameter depend on
the energy behavior of  the zeros
of the imaginary-part of the spin-flip amplitude
and the real-part of the spin-nonflip amplitude.
Figure 25 shows  $A_{N} (t)$ at $\sqrt{s}=9.2 $~GeV and $\sqrt{s}=13.7 $~GeV.
 At these energies the diffraction minimum deepens and its form affects  the form of $A_{N} (t)$.
 At last,  $A_{N} (t)$ is shown at large energies  $\sqrt{s}=19.4 $~GeV and $\sqrt{s}=23.4 $~GeV
 in Fig. 27.
 The diffraction dip in the differential cross section has a sharp form and it affects  the
 sharp form of $A_{N} (t)$.
The maximum negative values of $A_N$  coincide closely with
the diffraction minimum.   

 We have found that the contribution
of the spin-flip to the differential cross sections is much less
than the contribution of the spin-nonflip amplitude in the examined
region of momentum transfers. 
 $A_N$ is determined
in the domain of the diffraction dip by the ratio
\begin{eqnarray}
                 A_N \sim Im f_{- } / Re f_{+}. \label{ir}
\end{eqnarray}
The size of the analyzing power changes from $-45\%$ to $-50\%$
at $\sqrt{s}=50$ GeV up to $-25\%$ at $\sqrt{s}= 500$ GeV.
These numbers give the magnitude of the ratio, eq.(\ref{ir}), that does
not strongly depend on the phase between the spin-flip and
spin-nonflip amplitudes.  This picture implies that the diffraction minimum
is   mostly filled by the real-part of the spin-nonflip amplitude
and that the imaginary-part of the spin-flip amplitude
increases in this domain as well.

\section{Conclusions}


    Practically for the first time, a  simultaneous research of
    proton-proton,  proton-antiproton and proton-neutron elastic scattering
    has been carried out in a wide energy (from $3.6$ GeV up to $13$ TeV)
    and momentum transfer region (from $|t| = 2. 10^{-4}$ GeV$^2$ up to  $|t| = 14$ GeV$^2$).
  In the fitting procedure we used only statistical errors.
  Systematic errors, which are mostly determined by indefiniteness of luminosity,
  take into account as additional normalization coefficient.
   As a result, a wide range of possible forms of the scattering amplitudes
  decrease  considerably.
  In this paper
  a simultaneous  description of the  cross sections and spin correlation
   parameter of different nucleon-nucleon reactions, including 90 sets of experimental data,
   with the total number of data $N=4326$ gives a very reasonable $\sum_{i,j} \chi^2_{i,j} =4826$.
  The $pn$ case with $526$ experimental data, where the basic parameters were fixed from $pp$ and $p\bar{p}$ scattering,
   $\sum_{i,j} \chi^2_{i,j}=585$.

  Our analysis is carried out  by using 
 a successful development of the HEGS model which can be applied in a wide energy and momentum transfer regions.
   The model of hadron interaction is based on the analyticity of the scattering amplitude with taking into account the hadron structure, which is represented by GPDs.
  Different origins of the non-linear behavior of the slope of the scattering amplitude are analyzed.
  The possible contribution of a meson threshold is compared with different forms of the approximations
  for a non-linear slope at a small momentum transfer.

 The relative contributions of the possible different part of the scattering amplitude were especially analyzed.
  It is remarkable that in the model the main pomeron and odderon amplitudes
   have the same intercept.
  After eikonalization  this leads   to the $\ln^2(s)$ energy dependence.
   In this sense, we have some case of maximal Odderon.
    In the model, the odderon amplitude has a special kinematic factor and does not give a visible contribution at a zero momentum transfer.

  It was found that the new anomalous term with a large slope has the complicated logarithmic energy dependence and has the cross even properties. Hence, it is part of the pomeron amplitude and is also proportional to charge distributions. Our analysis of the contribution of the so-called hard pomeron with a large intercept does not show a visible contribution of this term.
The second additional term, which represents the  additional oscillation properties of the scattering amplitude
 at a small momentum transfer with the cross-odd properties,
has a logarithmic energy dependence and is proportional to the gravitomagnetic form factor.
 Hence, it belongs to the odderon contribution to the scattering amplitude.

  As was noted in many models, the oscillations of the scattering amplitude are connected with
  the broken  Pomeranchuk theorem  and the scattering amplitude grows to a maximal
possible extent but not breaking the Froissart boundary.
 In our opinion that oscillation of the scattering amplitude is connected with the behavior of hadron interaction potential
 at large distances \cite{Osc13,Sel-Uk,Sel-BogConf}.   Note that in \cite{1804.09564}  
 a quark-antiquark potential as function
of the separation distance $r$ between the particles for the
perturbative case in the framework of the Refined Gribov-Zwanziger Model (RGZ) 
 was calculated. It 
 includes a combination of oscillating terms.
 This can compared
to the Perturbative approach at short distances and Lattice QCD in the IR regime.
  So the researches into the properties of the oscillations term can give a very important information
  about the hadron interactions at large distances.
  Of course, this requires further theoretical and experimental works.

  This helps reduce
  some tension between the TOTEM and ATLAS data. 
 No contribution is shown of hard-Pomeron to  elastic hadron scattering. 
 However, the importance of Odderon's contribution is shown.
    A good description of proton-neutron differential scattering with 526 experimental points
  including the experimental data which reach extremely small momentum transfer $t=2 \ 10^{-4}$ GeV$^2$  is also obtained
  on the basis of the amplitudes taken from the analysis of
 $pp$ and $p\bar{p}$ scattering. A good  enough description of the polarization data was  also obtained,
 which reflects the true phases of the spin-non-flip and spin-flip amplitudes, so the value of
   $A_N(s,t) \sim \varphi_{sp-n-flip} - \varphi_{sp-flip}$.

   Our work supports that GPDs reflect the basic properties of the hadron structure
   and provide a bridge between  many different reactions.
   The determined new form of the momentum transfer dependence of GPDs
   allows one to obtain  different form factors, including Compton form factors,
   electromagnetic form factors, transition form factor, and gravitational form factors.
   The chosen form of the $t$-dependence of GPDs of the pion (the same as the $t$-dependence of the nucleon)
   allows us to describe the electromagnetic and gravitomagnetic form factors  of pion and pion-nucleon scattering.

    The impact of the different  $t$ dependence
   of the real and imaginary parts of the elastic scattering amplitude at small $t$
   should be noted.
   The dispersion relation shows that the real part has  zero at small $t$ , approximately in the
   domain $-t=0.1$; of course this depends on the energy.
    Hence, the contrary behavior of the real part (grows at small  $t$, the so-called "peripheral case"
   of the phase of the scattering amplitude) which is examined in \cite{TOTEM-8nexp} has no  physical meaning.

\vspace{0.5cm}
{\bf Acknowledgement}:
  {\small \hspace{0.3cm} OVS would  like to thank O. Teryaev and Yu. Uzikov
for their kind and helpful discussion.
This research was carried out at the expense of the grant of the Russian Science Foundation No. 23-22-00123, \\
 https://rscf.ru/project/23-22-00123. }\\


\end{document}